\documentclass[aps,pre,noshowkeys,
nosuperscriptaddress,
longbibliography,
reprint,
twocolumn, 
nofootinbib,
floatfix,
]{revtex4-1}

\usepackage[utf8]{inputenc}
\usepackage[OT4]{fontenc}
\usepackage{amsmath}
\usepackage{amssymb}
\usepackage{graphicx}
\usepackage{tikz}
\usepackage{subcaption}

\DeclareMathOperator{\Tr}{Tr}

\usepackage[colorlinks=true,hyperfootnotes=true,breaklinks=true]{hyperref}
\usepackage{cleveref}

\begin{document}
\title{Thermal properties of structurally balanced systems on diluted and densified triangulations}

\author{Maciej Wo{\l}oszyn}
\thanks{\includegraphics[width=10pt]{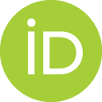}~\href{https://orcid.org/0000-0001-9896-1018}{0000-0001-9896-1018}}
\email{woloszyn@agh.edu.pl}
\affiliation{AGH University of Science and Technology,
Faculty of Physics and Applied Computer Science,
al. Mickiewicza 30, 30-059 Krak\'ow, Poland}

\author{Krzysztof Malarz}
\thanks{\includegraphics[width=10pt]{ORCID.png}~\href{https://orcid.org/0000-0001-9980-0363}{0000-0001-9980-0363}}
\email{malarz@agh.edu.pl}
\affiliation{AGH University of Science and Technology,
Faculty of Physics and Applied Computer Science,
al. Mickiewicza 30, 30-059 Krak\'ow, Poland}

\date{\today}

\begin{abstract}
    The dynamics of social relations and the possibility of reaching the state of structural balance (Heider balance) are discussed for various  networks of interacting actors under the influence of the temperature modeling the social noise level.
    For that purpose, two main types of lattices are considered. 
    The first is created by removing some links from a regular triangular lattice to produce a diluted triangular lattice, and the second by adding more links to create an enhanced triangular lattice.
    In both those cases, the full range of possible graph densities is discussed, limited by the extreme cases of networks which consist of a small number of separated triads and fully connected networks. 
    It is shown that the existence of the balanced state is not possible
    if the average node degree is too close to the value characterizing the regular triangular lattice.
    Otherwise, both balanced (or partially balanced) and imbalanced states are possible, depending on the temperature.
    However, only for graphs which are dense enough a phase transition of the first kind is observed, while less enhanced networks (and all diluted) indicate a smooth cross-over between the two states.
    The cross-over temperatures are size-independent only for the diluted triangular lattices and depend on the size of the system for the enhanced triangular lattices, as is the case also for the critical temperatures of the phase transition observed in denser enhanced lattices.
\end{abstract}
\maketitle

\section{Introduction}

The importance of social processes has been the reason of broad interest in their modeling with methods of statistical physics.
One of the most interesting problems in that field is the evolution of social networks towards the state known as the Heider (or structural) balance \cite{Heider,Harary_1953,Cartwright_1956,Harary_1959,Davis_1967,Harary}.
Dynamics of such networks has been studied on various underlying lattices, from chains of actors \cite{2008.06362} to complete graphs \cite{Antal_2005}, specifically to find whether vanishing of the ordered and balanced phases is possible and under what conditions.
Reviews of the discussed problems and used methods are given in Refs.~\onlinecite{ISI:000247470200019,Belaza_2017}.
Recent examples of the application of the balance theory include 
usage of the competitive balance model with two different interests \cite{2008.00537},
the coevolutionary balance model in which actors change both their opinions and relationships \cite{2010.10036},
the study of social fragmentation and its influence on the dynamics of opinion formation \cite{Pham_2005.01815}, 
or analysis of indirect reciprocity and its impact on the friendship and enmity relations in the network \cite{2104.10568}. 
In all those cases, the model applied to perform either calculations or numerical simulations is based on Heider's concept of triads of actors, and friendly or hostile relations between those actors.

\begin{figure}[htbp!]
\begin{subfigure}{.240\columnwidth}
\centering
\caption{\label{fig:1a}}
\begin{tikzpicture}[scale=1.0]
\draw[blue,very thick] (0,0) -- (1,0);
\node[below] at (0.5, 0) {$+$};
\draw[blue,very thick] (0,0) -- (0.5,0.866);
\node[left] at (0.25, 0.433) {$+$};
\draw[blue,very thick] (1,0) -- (0.5,0.866);
\node[right] at (0.75, 0.433) {$+$};
\end{tikzpicture}
$s=+3$
\end{subfigure}
\begin{subfigure}{.240\columnwidth}
\centering
\caption{\label{fig:1b}}
\begin{tikzpicture}[scale=1.0]
\draw[blue,very thick] (2,0) -- (3,0);
\node[below] at (2.5, 0) {$+$};
\draw[blue,very thick] (2,0) -- (2.5,0.866);
\node[left] at (2.25, 0.433) {$+$};
\draw[red,very thick, dashed]  (3,0) -- (2.5,0.866);
\node[right] at (2.75, 0.433) {$-$};
\end{tikzpicture}
$s=+1$
\end{subfigure}
\begin{subfigure}{.240\columnwidth}
\centering
\caption{\label{fig:1c}}
\begin{tikzpicture}[scale=1.0]
\draw[blue,very thick] (4,0) -- (5,0);
\node[below] at (4.5, 0) {$+$};
\draw[red,very thick, dashed]  (4,0) -- (4.5,0.866);
\node[left] at (4.25, 0.433) {$-$};
\draw[red,very thick, dashed]  (5,0) -- (4.5,0.866);
\node[right] at (4.75, 0.433) {$-$};
\end{tikzpicture}
$s=-1$
\end{subfigure}
\begin{subfigure}{.240\columnwidth}
\centering
\caption{\label{fig:1d}}
\begin{tikzpicture}[scale=1.0]
\draw[red,very thick, dashed]  (6,0) -- (7,0);
\node[below] at (6.5, 0) {$-$};
\draw[red,very thick, dashed]  (6,0) -- (6.5,0.866);
\node[left] at (6.25, 0.433) {$-$};
\draw[red,very thick, dashed]  (7,0) -- (6.5,0.866);
\node[right] at (6.75, 0.433) {$-$};
\end{tikzpicture}
$s=-3$
\end{subfigure}
\caption{\label{fig:triads}(Color online). Heider's triads corresponding to balanced (\ref{fig:1a} and \ref{fig:1c}) and imbalanced (\ref{fig:1b} and \ref{fig:1d}) states.
Solid blue lines and dashed red lines represent friendly ($+1$) and hostile ($-1$) relations, respectively.}
\end{figure}
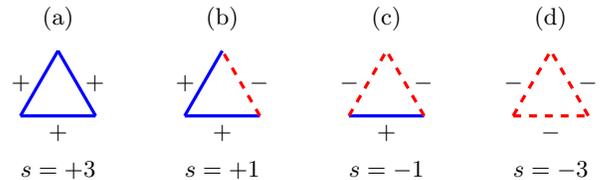

The system contains $N$ labeled ($1\le i\le N$) actors (nodes) and $L$ relations (links) among these actors.
The $+1$ ($-1$) link values $x_{ij}$ indicate friendly (hostile) relations among the actors $i$ and $j$.
The system dynamics is governed by the changes of links values in actor triangles.
There are four available types of triangles in this system, as presented in \Cref{fig:triads} with $s$ being equal to the sum of link values in the given triad.

Among them, two (those presented on \Cref{fig:1a,fig:1c}) are termed as balanced (in Heider's sense) as the signs of relations obey the following rules:
\begin{itemize}
\item a friend of my friend is my friend,
\item a friend of my enemy is my enemy,
\item an enemy of my friend is my enemy,
\item an enemy of my enemy is my friend.
\end{itemize}
In the imbalanced triangles (\Cref{fig:1b,fig:1d}) these rules are violated what leads to the appearance of mental stress known as cognitive dissonance.
Resolving this stress may be achieved by changing some of the links values in the imbalanced triangle, which is the source of the system temporal evolution towards the Heider balance, where the triangles \ref{fig:1b} and \ref{fig:1d} are absent.

Most of the earlier efforts in Heider balance research were devoted to the study of the structural balance on complete graphs \cite{Antal_2005} as the assumed graph geometry ensured both: 
\begin{itemize}
\item presence of triads necessary for introducing cognitive dissonance among actors 
\item and simultaneously allows for some analytical considerations including mean-field calculations \cite{PhysRevE.99.062302,1911.13048}.
\end{itemize}

Recently, the systems were enriched with thermal noise modeled either with Glauber dynamics \cite{PhysRevE.100.022303,2009.10136,2011.07501} or heat-bath \cite{PhysRevE.99.062302,1911.13048}. 
The results of Refs.~\onlinecite{PhysRevE.100.022303,2009.10136,PhysRevE.99.062302,1911.13048} indicate that a phase transition occurs in the system: below the critical noise level $T<T_C$ the system orders into the balanced (in Heider's sense) state, while for the high enough noise level $T>T_C$ the system is imbalanced. The balance/imbalance phase transition was identified as the first-order phase transition. The critical temperature $T_C$ increases linearly with the system size $N$.

On the other hand, the simplest lattice (where triangles appear in a natural way) is the triangular lattice. 
The structural balance on such network was studied with (deterministic) cellular automata \cite{2005.11402} and with the (stochastic) heat-bath algorithm \cite{2007.02128}.
The latter study showed that independently of the assumed noise level, the system reaches the imbalanced state. In other words, the critical temperature in such a system tends to zero ($T_C\to 0$).

Finally, the Heider balance may be observed also in a chain of actors where triangles may be introduced by adding long-range interactions \cite{2008.06362}. There, depending on the interaction range, the signatures of absence (for the nearest neighbor interactions assumed), the second-order (for the intermediate range of interaction), and the first-order phase transition (for a very long range of interactions) were reported. 
The critical temperature $T_C$ increases non-linearly with the interaction range and the initial fraction of friendly relations.

Intrigued by the above-mentioned (induced by underlying network connectivity) variety of system complex behaviors, here we systematically check how the network topology influences the Heider balance in the presence of (thermal) noise. 
To that end, we enrich (towards a complete graph) or dilute (towards separated triangles) the triangular lattice and apply the heat-bath algorithm to simulate the evolution of links.

We note that more intricate behavior may be observed after further enrichment of the system with additional interactions based not only on link values but also on attributes assigned to the network nodes \cite{PhysRevLett.125.078302,Pham_2005.01815,2010.10036}.

\section{Model}

The system Hamiltonian for a fully connected graph \cite{Antal_2005,PhysRevE.99.062302},
\begin{equation}
\mathcal{H}=-\frac{1}{6}\Tr(\mathbf{X}^3), 
\end{equation}
where $\mathbf{X}=[x_{ij}]$,
may be easily adopted for any network with $N$ nodes and $L$ links.
In such case it becomes
\begin{equation}
\mathcal{H}=- \frac{1}{6} \Tr[(\mathbf{A}\circ\mathbf{X})^3], 
\end{equation}
where (binary and symmetric) adjacency matrix $\mathbf{A}=[a_{ij}]$ elements
\begin{equation}
a_{ij}=\begin{cases}
1 & \iff i \text{ and } j \text{ are connected},\\
0 & \iff \text{otherwise},
\end{cases}
\end{equation}
define the network, $\circ$ stands for the Hadamard product of matrices, and $x_{ij}=0$ when there is no link among the nodes $i$ and $j$ ($a_{ij}=0$).
For a complete graph, $a_{i,i}=0$ and $a_{i,j\ne i}=1$.

\begin{figure}[htbp!]
\begin{subfigure}[b]{.39\textwidth}
\caption{\label{fig:A_T_a}}
\includegraphics[width=0.99\textwidth]{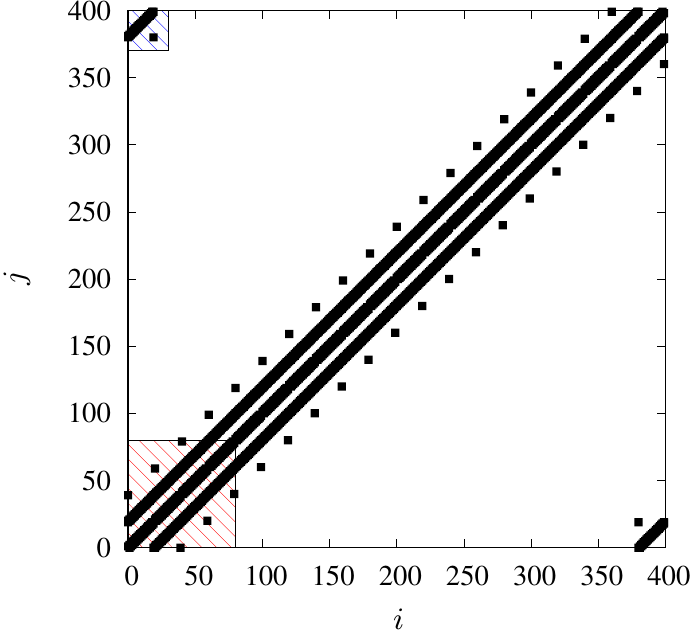}
\end{subfigure}
\begin{subfigure}[b]{.39\textwidth}
\caption{\label{fig:A_T_b}}
\includegraphics[width=0.99\textwidth]{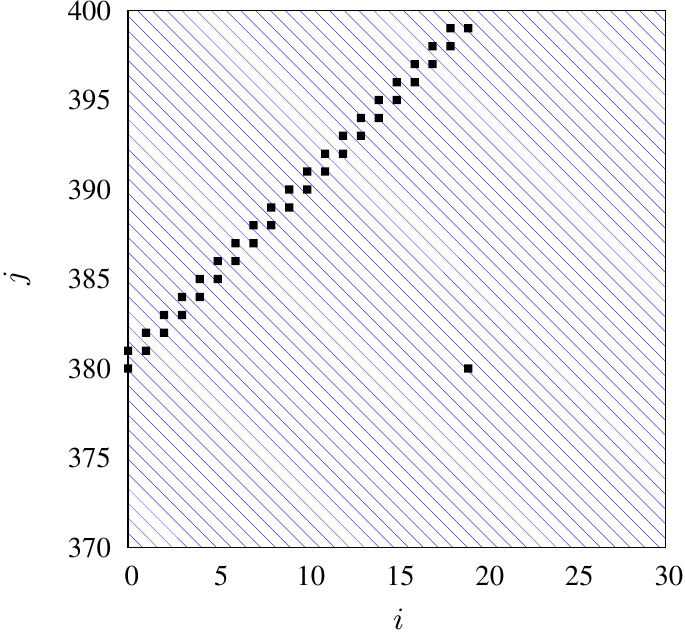}
\end{subfigure}
\begin{subfigure}[b]{.39\textwidth}
\caption{\label{fig:A_T_c}}
\includegraphics[width=0.99\textwidth]{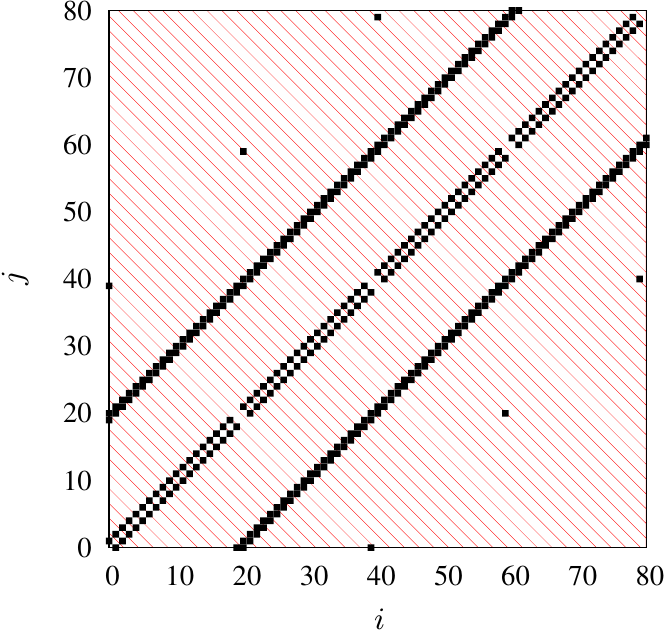}
\end{subfigure}
\caption{\label{fig:A_T}(Color online). Shape of the adjacency matrix $\mathbf{A}=[a_{ij}]$ for the triangular lattice on torus for $W=20$. $a_{ij}=1$ values are marked by black squares. (a) Whole matrix ($1\le i,j\le W^2$), (b) upper-left matrix corner, (c) bottom-left matrix corner.} 
\end{figure}

A triangular lattice may be constructed based on a square grid with $N=W^2$ sites, where $W$ is the linear size of the lattice.
Then the neighbors of site $(n,m)$, $1\le n,m\le W$, are located at $(n-1,m)$, $(n+1,m)$, $(n,m-1)$, $(n,m+1)$, $(n+1,m-1)$, $(n-1,m+1)$. We assume periodic boundary conditions.
The adjacency matrix $\mathbf{A}$ for the triangular lattice with periodic boundary conditions is schematically sketched in \Cref{fig:A_T}.

To construct the diluted triangular lattice (DTL), a fraction of the unit elements ($a_{i,j}=a_{j,i}=1$) must be replaced with zeros ($a_{i,j}=a_{j,i}=0$). 
We implement this dilution by removing exactly $q_-$ randomly chosen links.
Similarly, adding extra links between the nodes of the triangular lattice to create the enhanced triangular lattice (ETL) requires substituting $0\to 1$ in the adjacency matrix and keeping it symmetric.
It means that adding $q_+$ links is performed by replacing $q_+$ randomly chosen pairs of zero elements ($a_{ij}=a_{ji}=0$) by unit values ($a_{ij}=a_{ji}=1$).

Having the adjacency matrix $\mathbf{A}$, the number of links is
\begin{equation}
L=\frac{1}{2}\Tr(\mathbf{A}^2)
\end{equation}
and the number of triads may be calculated as
\begin{equation}
\Delta=\frac{1}{6}\Tr(\mathbf{A}^3).
\end{equation}
In the limiting cases, these values are known analytically: for a complete graph 
$L_{\textsc{cg}}=\binom{N}{2}$ and $\Delta_{\textsc{cg}}=\binom{N}{3}$,
while for a triangular lattice with periodic boundary conditions
$L_{\textsc{tr}}=3N=3W^2$ and $\Delta_{\textsc{tr}}=2N=2W^2$.

For the diluted systems, removing $q_-$ links yields the bond occupation probability $p_-=1-q_-/L_{\textsc{tr}}$ which seems to be a natural measure of DTL connectivity.
Graphs corresponding to those lattices have the $p_-$ parameter whose value is between $p_-=0$ (an empty graph) and $p_-=1$ (a triangular lattice with periodic boundary conditions).
Enriching the triangular lattice leads to the change of its structure in the opposite direction, starting from $p_+=1$ towards a larger number of links.
By analogy to $p_-$, the occupation parameter $p_+$ is defined as $p_+=1+q_+/L_{\textsc{tr}}$, where $q_+$ is the number of added links.
It describes how many times the number of links is larger compared to the triangular lattice from which ETL was constructed.
As both $p_-=1$ and $p_+=1$ correspond to the regular triangular lattice, for each of them the average node degree is $\bar{k} = 6$, and in general $\bar{k} = 6 p_\pm$.

Since for a complete graph $L_{\textsc{cg}}=\binom{N}{2}$, the graph density is
\begin{equation}
D= \frac{L}{L_{\textsc{cg}}} = \frac{\Tr(\mathbf{A}^2)}{2\binom{N}{2}}.
\end{equation}
It can be related to the bond occupation probability if we note that the number of links removed from the triangular lattice to create its diluted version is $q_-=L_{\textsc{tr}}-L$, or $q_+=L-L_{\textsc{tr}}$ when we enhance the triangular lattice with $q_+$ more links, where $L$ is the number of links in the resulting diluted or enriched lattice.
For both of them, it leads to $p_\pm=L/L_{\textsc{tr}}=D/D_{\textsc{tr}}$. 
Because $D_{\textsc{tr}}=L_{\textsc{tr}}/L_{\textsc{cg}}=6/(N-1)$, it follows that the graph density depends on the bond occupation probability as $D=6p_\pm/(N-1)=\bar{k}/(W^2-1)$.
For example, the density of the triangular lattice with linear size $W=10$ is $D_{\textsc{tr}} \approx 0.061$, while for $W=20$ it is $D_{\textsc{tr}} \approx 0.015$.

The Heider balance may be easily identified by checking the system work function \cite{Antal_2005,Krawczyk_2017}
\begin{equation}
\label{eq:Udef}
U\equiv-\frac{\Tr[(\mathbf{A}\circ\mathbf{X})^3]}{\Tr(\mathbf{A}^3)}.
\end{equation}
The system work function $U$ is equal to $-1$ if and only if all triangles in the system are balanced.

We assume that the system evolution is governed by the heat-bath algorithm \cite{PhysRevB.33.7861,Loison_2004}. 
If $a_{ij}\ne 0$ then evolution of the link between nodes $i$ and $j$ of value $x_{ij}$ is given by
\begin{subequations}
\label{eq:evol}
\begin{equation}
\label{eq:evol_x}
x_{ij}(t+1)=
        \begin{cases}
	+1 & \text{ with probability }p_{ij}(t),\\
	-1 & \text{ with probability }[1-p_{ij}(t)],
        \end{cases}
\end{equation}
where
\begin{equation}
\label{eq:evol_p}
    p_{ij}(t)=\frac{\exp[\xi_{ij}(t)/T]}{\exp[\xi_{ij}(t)/T]+\exp[-\xi_{ij}(t)/T]}.
\end{equation}
$T$ is the temperature (noise level) at which the evolution occurs, and
\begin{equation}
\label{eq:evol_xi}
	\xi_{ij}(t)=\sum_{k} a_{ik} x_{ik}(t) \cdot a_{kj} x_{kj}(t).
\end{equation}
\end{subequations}
This procedure allows us to carry out the stochastic evolution of the system.
It means that in addition to the tendency to minimize the work function $U$ by changing the state of triads from imbalanced to balanced, which modifies the state of the system towards the limit of $U \to -1$,  also the changes increasing $U$ are possible with non-zero probability given by \Cref{eq:evol}.
As a result, the system evolves towards the thermal equilibrium realized at some $U > -1$.
In each time step of the evolution, all links are updated synchronously.

\section{Results}

Simulations based on the models described in the previous section were performed to find under what conditions the system changes its state from balanced to imbalanced, and whether a phase transition or a smooth cross-over between those states occurs.
Since the main objective is to identify the values of the characteristic temperature $T_C$ at which such transition or cross-over is observed, we (quite arbitrarily) assume that $T=T_C$ when $U(T)=-0.5$ to be consistent with Refs.~\onlinecite{2008.06362,2009.10136}.
The characteristic temperature identified in such a manner is denoted as $T_C^U$ from now.

Below we discuss our results obtained for systems of two different types:  when links are gradually randomly removed from a triangular lattice, thus reducing the graph density (\Cref{sec:DTL}) and when the underlying lattice is created by adding links to a triangular lattice up to the point when it becomes a complete graph (\Cref{sec:ETL}).
In each of those cases, we examine the full range of the possible densities of the considered graphs.

We start simulation with the system in the imbalanced ($U=0$) initial state achieved by setting the same number of positive and negative link values at $t=0$.
The simulations are carried out for systems with $N=W^2$ sites and $W=10$, $15$ or $20$.

\subsection{\label{sec:DTL}Diluted triangular lattice}

As it was shown in Ref.~\onlinecite{2007.02128}, for the triangular lattice ($q_-=0$, $p_-=1$) the system reaches the imbalanced state independently on the assumed temperature $T$.

We note that the system work function $U$ is multiply degenerated as dozens among $2^L$ available positive/negative link distributions may lead to the same value of $U$.
The triangular lattice dilution ($q_->0$, $p_-<1$) introduces defects in the system which reduce the number of available work function $U$ values.
For high temperature $T$, independently on the system dilution $q_-$, the work function $U$ fluctuates around $U=0$.
For $T=0.5$ and $p_-=1/2$ the work function fluctuates around $-0.7$ and for further lattice dilution ($p_-=1/5$) it switches mainly between the four discrete, well separated but negative values of $-1\le U<0$ ($U\in\{-1, -0.714, -0.428, -0.143\}$).

\begin{figure}[htbp]
\begin{subfigure}[b]{.48\textwidth}
\caption{\label{fig:DTL_U_vs_T_W10}}
\includegraphics[width=0.99\textwidth]{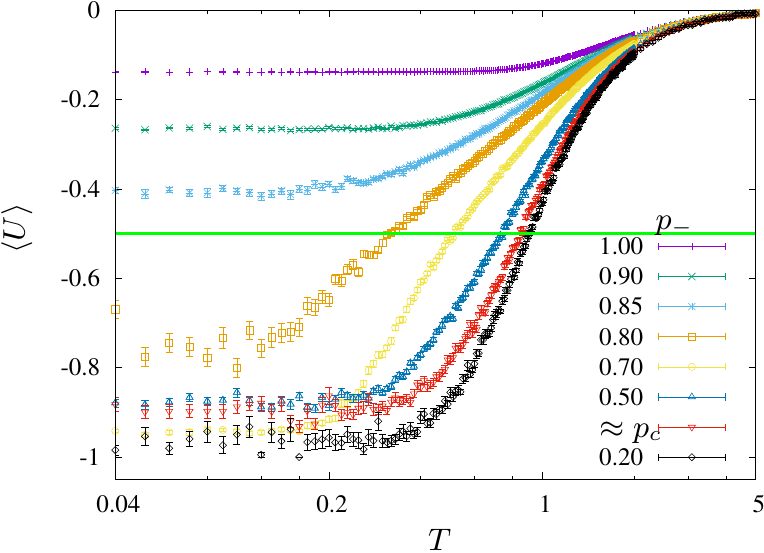}
\end{subfigure}
\begin{subfigure}[b]{.48\textwidth}
\caption{\label{fig:DTL_U_vs_T_W20}}
\includegraphics[width=0.99\textwidth]{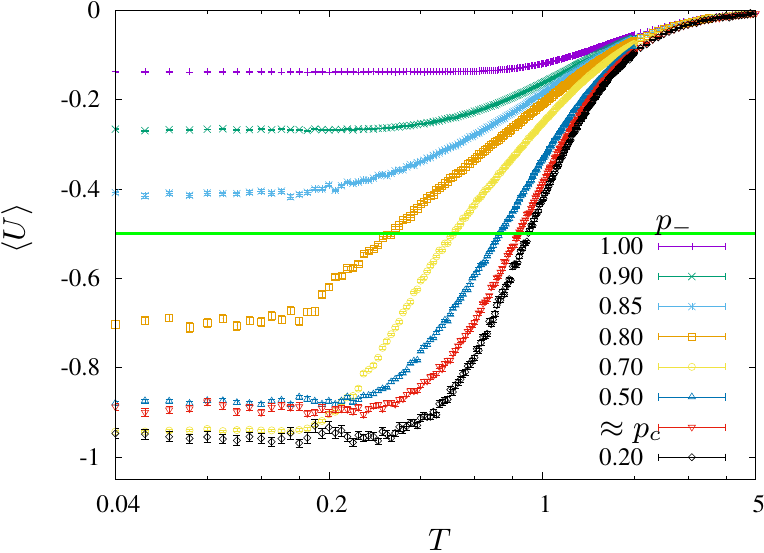}
\end{subfigure}
\caption{\label{fig:DTL_U_vs_T}(Color online). Thermal evolution of work function $\langle U\rangle$ for diluted triangular lattices, $t_{\max}=10^4$, $\tau=10^3$, $R=100$, various $p_-$,  (a) $W=10$, (b) $W=20$.
The bond percolation threshold for the triangular lattice $p_c=2\sin(\pi/18)\approx 0.347\cdots$.}
\end{figure}

In \Cref{fig:DTL_U_vs_T} the thermal evolution of the average work function $\langle U\rangle$ for  various values of the system dilution (expressed in terms of the bond occupation probability $p_-$) and various lattice sizes ($W=10$ and 20) is presented. 
A single point in this chart comes from averaging over the last $\tau=10^3$ among $t_\text{max}=10^4$ time steps of the evolution, and
the uncertainty of the point position comes from averaging over $R=100$ simulations.
The averaging procedure over both last time steps and various independent simulations is denoted as $\langle\cdots\rangle$.
The system size $N=W^2$ does not influence the results significantly.
In this aspect (i.e., independence of the results on the system size $N$) the diluted triangular lattice is similar to the one-dimensional chain of nodes/links with long-range interactions \cite{2008.06362}. 
However, this is quite different than, for instance, for a complete graph where $T_C$ increases with system size as $T_C\propto (N-2)$ \cite{1911.13048}.

\begin{figure}[htbp]
\begin{subfigure}[b]{.48\textwidth}
\caption{\label{fig:UpT_rho050_W10}}
\includegraphics[width=0.99\textwidth]{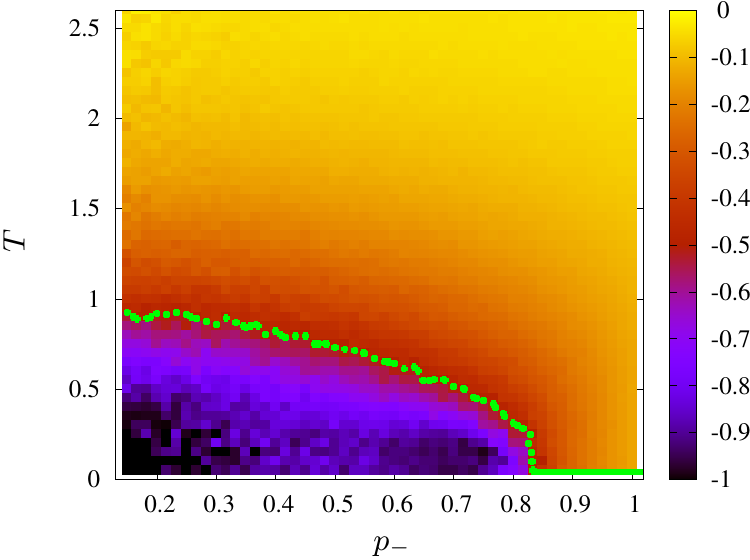}
\end{subfigure}
\begin{subfigure}[b]{.48\textwidth}
\caption{\label{fig:UpT_rho050_W20}}
\includegraphics[width=0.99\textwidth]{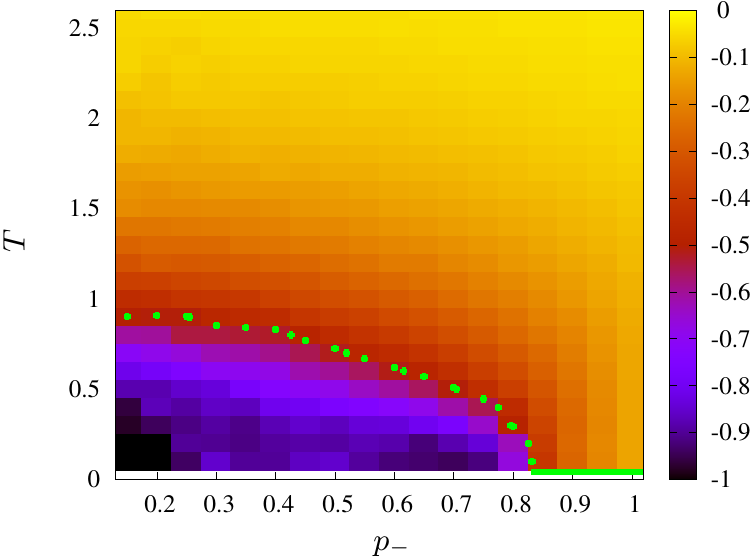}
\end{subfigure}
\caption{\label{fig:DTL_U_vs_p_T}(Color online) Average work function $\langle U\rangle$ vs. occupation probability $p_-$ and temperature $T$ for DTL, $R=10$, $t_{\max}=10^4$, $\tau=10^3$, and (a)~$W=10$ or (b)~$W=20$. Green dots indicate the points $(p_-,T)$ where $U=-0.5$ and thus at this border we have $T=T_C^U$.}
\end{figure}

As for the dilution level in the range $0.83<p_-<1$ where the system stays closer to the imbalanced state than to the balanced one (and the curve $U(T)$ does not intersect the line $U=-0.5$), we may assume that in this dilution region the characteristic temperature tends to zero ($T_C^U\to 0$) what is consistent with our earlier studies \cite{2007.02128}.
Finally, independently on  the lattice dilution $p_-$, the system reaches the imbalanced ($U=0$) for high enough noise level $T>5$.

In \Cref{fig:DTL_U_vs_p_T} the dependencies of $\langle U\rangle$
vs. the occupation probability $p_-$ and thermal noise $T$ for $W=10$ and $W=20$  are presented, with boundary corresponding to $U=-0.5$ (i.e., the characteristic temperature $T_C^U(p_-)$ dependence on the occupation probability $p_-$) is marked with green points.
The results are averaged over $R=10$ simulations.

To sum up the results obtained for DTL: 
\begin{itemize}
\item the triangular lattice dilution highlights the degeneration of the system work function $U$; 
\item if only a small number of links is removed from the triangular lattice ($0.83<p_-\le 1$) the system never reaches $U<-0.5$ (see \Cref{fig:DTL_U_vs_T,fig:DTL_U_vs_p_T}), which may be considered as  $T_C^U\to 0$; 
\item the average work function $\langle U\rangle$ depends on the triangular network dilution $q_-$ and the assumed noise level $T$ (see \Cref{fig:DTL_U_vs_p_T});
\item reaching in dilution the bond percolation threshold and further ($p_-<p_c=2\sin(\pi/18)\approx 0.347\cdots$ \cite{bookDS}) does not change the system behavior qualitatively, and for $p_-\to 0$ we have $T_C^U\to 1$;
\item and the system size $N=W^2$ does not influence the results.
\end{itemize}

\subsection{\label{sec:ETL}Enhanced triangular lattice}

Simulations on the lattices created by adding links to the initial triangular lattice were performed as in the previous case, starting from a random imbalanced ($U=0$) initial state with fifty-fifty distribution of friendly and hostile relations.
Same as for the diluted triangular lattice ($p_-<1$), the imbalanced state is retained at higher temperatures for all considered graph densities, however the required temperatures tend to be larger as $p_+$ increases.
On the other hand, no quantization of the work function values is visible, because adding links does not lead to reduction of the available values of $U$; on the contrary, the number of possible work function values increases and the spectrum remains quasi-continuous.

\begin{figure}[htbp]
\begin{subfigure}[b]{.48\textwidth}
\caption{\label{fig:ETL_U_vs_T_W10}}
\includegraphics[width=0.99\textwidth]{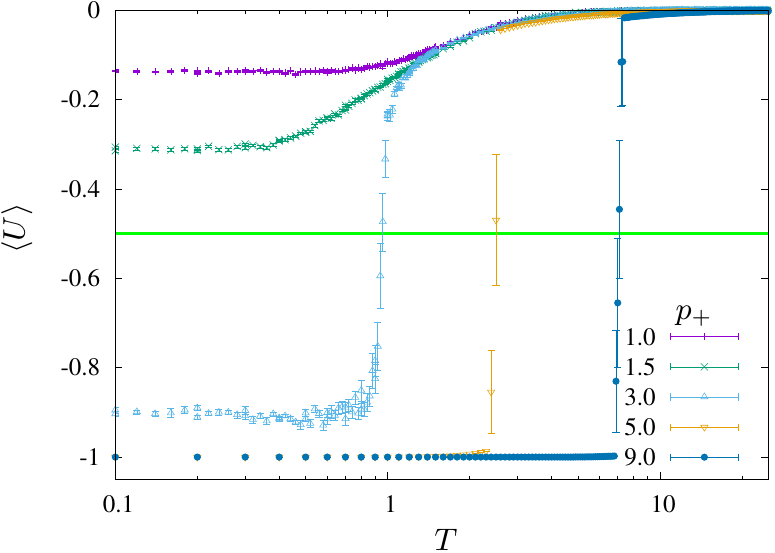}
\end{subfigure}
\begin{subfigure}[b]{.48\textwidth}
\caption{\label{fig:ETL_U_vs_T_W20}}
\includegraphics[width=0.99\textwidth]{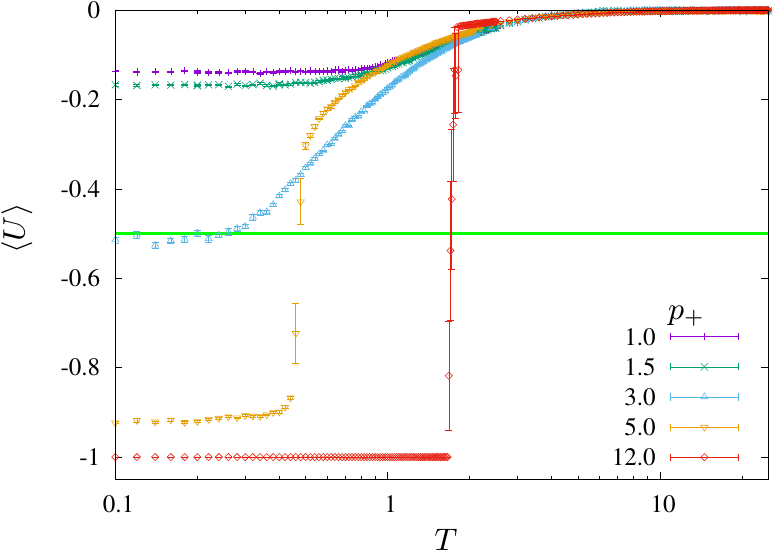}
\end{subfigure}
\caption{\label{fig:ETL_U_vs_T}(Color online). Thermal evolution of average work function $\langle U\rangle$ for ETL, $t_{\max}=10^4$, $\tau=10^3$, $R=10$, various $p_+$, (a) $W=10$, (b) $W=20$.}
\end{figure}

The average value of the work function $\langle U\rangle$ is presented as a function of temperature in \Cref{fig:ETL_U_vs_T} for $W=10$, $20$ and several values of $p_+$, starting from $p_+=1$ which corresponds to the triangular lattice.
All points were obtained from the averaging procedure over the last $\tau = 10^3$ steps of $R=10$ simulations which took $t_\text{max}=10^4$ time steps each.
Unlike for $p_-<1$, those characteristics strongly depend on the system size $N=W^2$.
Comparison of \Cref{fig:ETL_U_vs_T_W10,fig:ETL_U_vs_T_W20} (e.g. for $p_+=5$) reveals that for the smaller size $W$ the balanced phase can exist also at temperatures which are higher than for larger systems with the same $p_+$.

\begin{figure}[htbp]
\begin{subfigure}[b]{.48\textwidth}
\caption{\label{fig:ETL_U_vs_p_T_W10}}
\includegraphics[width=0.99\textwidth]{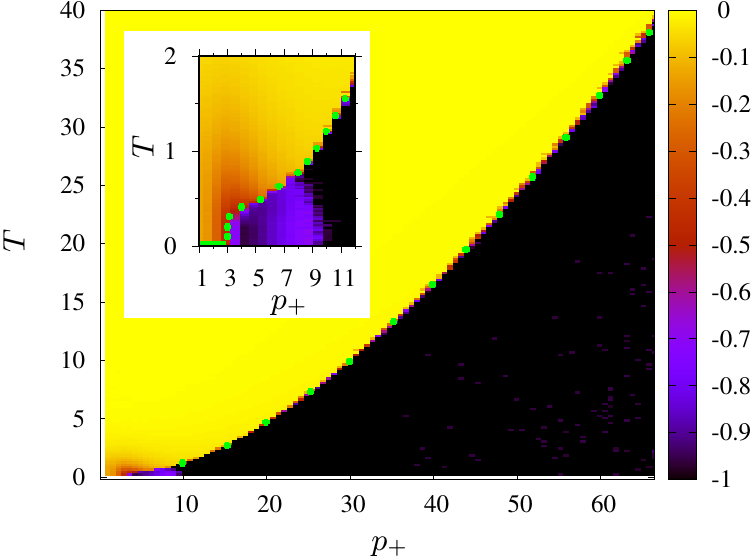}
\end{subfigure}
\begin{subfigure}[b]{.48\textwidth}
\caption{\label{fig:ETL_U_vs_p_T_W20}}
\includegraphics[width=0.99\textwidth]{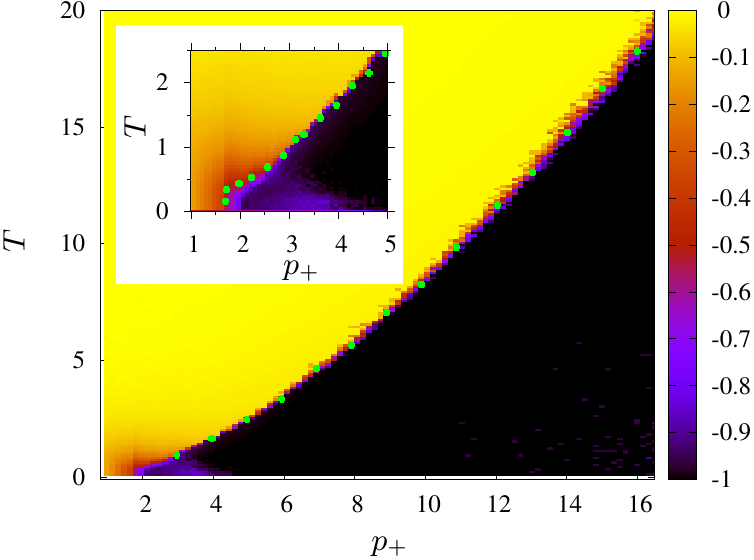}
\end{subfigure}
\caption{\label{fig:ETL_U_vs_p_T}(Color online) Average work function $\langle U\rangle$ vs. occupation parameter $p_+$ and temperature $T$ for ETL, $R=10$, $t_{\max}=10^4$, $\tau=10^3$, and (a) $W=10$ or (b) $W=20$. Green dots indicate the points $(p_+,T)$ where $U=-0.5$ and thus at this border we have $T=T_C^U$.}
\end{figure}

The dependence of the average value of the work function $\langle U\rangle$ on the occupation parameter $p_+$ and the temperature $T$ is presented in \Cref{fig:ETL_U_vs_p_T} for $W=10$ and $W=20$. 
It shows that the boundary between the regions of the $(p_+,T)$-space with balanced and imbalanced phases is very sharp, except for the smallest values of $p_+$ (see the insets of \Cref{fig:ETL_U_vs_p_T}).
This feature is also visible in the $U(T)$ dependencies presented in 
\Cref{fig:ETL_U_vs_T} (for $p_+ \geq 5$).

In contrast to DTL, in the case of ETL the characteristic temperatures are not size-independent.
Their values are indicated by the boundary (green dots) in \Cref{fig:ETL_U_vs_p_T}, following the condition $U=-0.5$.
The maximum value of the bond occupation parameter, $p_{\text{max}}$, for which the results are presented is different in \Cref{fig:ETL_U_vs_p_T_W10} and in \Cref{fig:ETL_U_vs_p_T_W20} which stems from the fact that the limiting case is always the complete graph.
Then, the density is $D=1$ and the corresponding $p_{\text{max}}=(W^2-1)/6$ depends on the size of the system; e.g. for $W=10$ we have $p_{\text{max}}=16.5$ while for $W=20$ it is equal $p_{\text{max}}=66.5$.
At those values of the bond occupation parameter, the lattice reaches the limit of the complete graph for which the critical temperatures are known to be size-dependent~\cite{2009.10136} and are in agreement with our results for $p_+=p_{\text{max}}$.

Cross-over from an imbalanced state to a balanced state is observed at a positive $T_C^U$ only if the number of links added to the triangular lattice is above some minimum value, and hence the bond occupation parameter is above the corresponding minimum value $p^{*} > 1$.
Bearing in mind that for DTL discussed in \Cref{sec:DTL} $T_C^U>0$ was found only if the number of removed links was large enough, specifically when $p_-<0.83$, it means that $T_C^U\to 0$ for a certain range of $p$-parameter values around $p_\pm=1$ (which corresponds to the triangular lattice).
It proves that the underlying network geometry has to be considerably different from the triangular lattice to produce $T_C^U>0$ and allow for a balanced state.
However, there is an important difference between the situations for ETL ($p_+>1$) and DTL ($p_-<1$).
Unlike in the latter case, the system size affects the value of $p_+=p^*$ which separates the cases where the finite and positive characteristic temperatures can be found from those with $T_C^U \to 0$, e.g. for $W=10$ it is $p^*\approx 1.7$ (see the inset in \Cref{fig:ETL_U_vs_p_T_W10}) and increases to $p^* \approx 2.9$ for $W=20$ (see the inset in \Cref{fig:ETL_U_vs_p_T_W20}).

The main results for ETL are in short:
\begin{itemize}
    \item if the number of links added to the triangular lattice is small, $p_+ < p^*$ (e.g., $p_+ < 1.7$ for $W=10$ or $p_+ < 2.9$ for $W=20$), the system cannot be even partially balanced and $U > -0.5$ (see \Cref{fig:ETL_U_vs_T,fig:ETL_U_vs_p_T});
    \item the average work function $\langle U\rangle$, similarly as in DTL, depends on the network enhancement parameter and the noise level, but the boundary between the imbalanced and (at least partially) balanced states becomes very sharp when $p_+$ increases;
    \item the characteristic temperature and the interval of $p_+$ in which it is defined depend on the system size. 
\end{itemize}

\section{Discussion}

\begin{figure}[htbp]
\begin{subfigure}[b]{.48\textwidth}
\caption{\label{fig:all-TC_vs_p-a}}
\includegraphics[width=0.99\columnwidth]{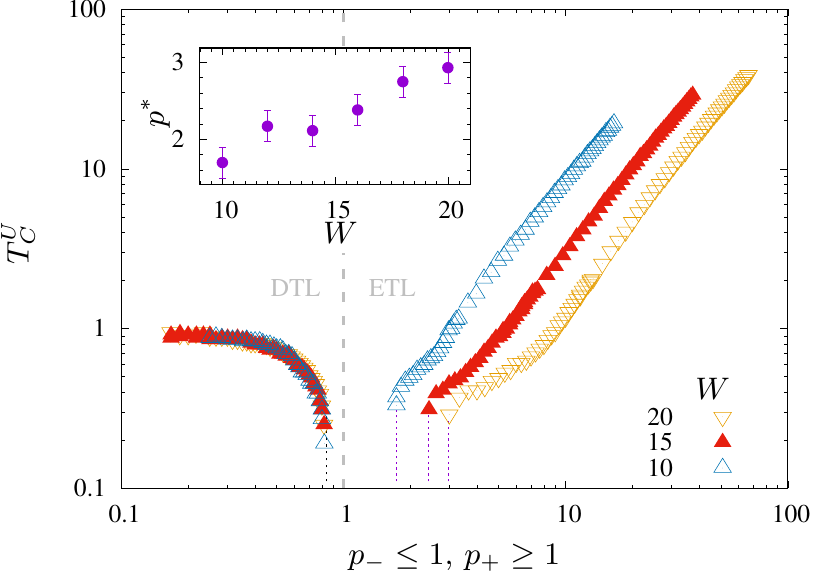}
\end{subfigure}
\begin{subfigure}[b]{.48\textwidth}
\caption{\label{fig:all-TC_vs_p-b}}
\includegraphics[width=0.99\columnwidth]{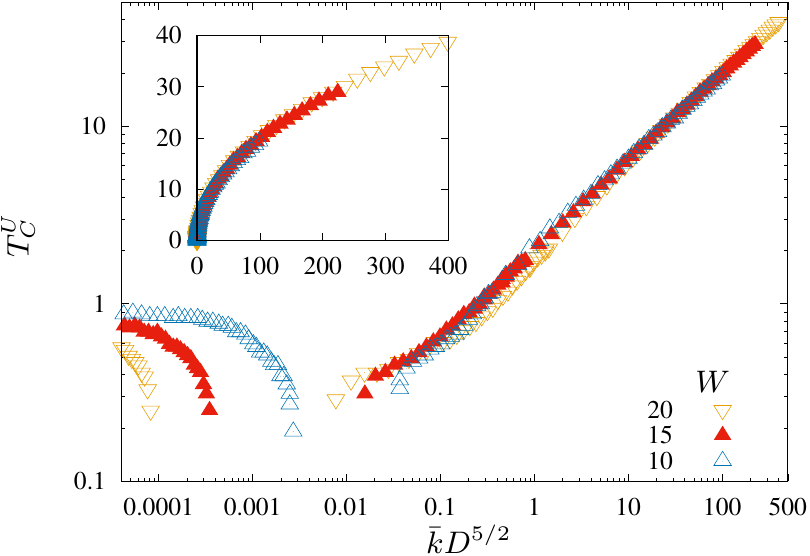}
\end{subfigure}
\caption{\label{fig:all-TC_vs_p}(Color online). (a) Characteristic temperature $T_C^U$ vs. occupation parameter $p_\pm$ for DTL (when $p_-\le 1$) or ETL (when $p_+\ge 1$).
Inset: the minimum value $p^*$ (also indicated with the dashed lines in the main part of this figure) required for $T_C^U > 0$ in ETL depending on its size $W$.
(b) The same data presented vs. $\bar kD^{5/2}$, which leads to data collapse into single curve for a complete graph limit. In the inset, linear scale is used instead of logarithmic.
Both (a) and (b) were obtained with $R=10$, $t_{\text{max}}=10^4$, $\tau=10^3$.}
\end{figure}

The characteristic temperatures found from the condition $U=-0.5$ for both DTL and ETL are presented in \Cref{fig:all-TC_vs_p} where the rightmost points for each $W$ correspond to complete graphs, i.e., $D=1$.
Analysis of the lattices with $p_-<1$ reveals that for DTL they approach the limiting value $T_C^U=1$ when $p_- \to 0$.
It is not surprising since the reduction of the graph density inevitably leads towards systems consisting of a small number of disconnected triads.
When $p_-$ increases, a gradual decrease in $T_C^U$ is visible, and  $T_C^U \to 0$ at $p_-=5/6$. 
Then, the characteristic temperature remains undefined until the lattice not only changes into ETL with $p_+>1$ but also $p_+$ reaches the size-dependent minimum value $p^{*}$ which  increases with the size of the system (see the inset in \Cref{fig:all-TC_vs_p-a} presenting $p^{*}$ for $W=10, 12, \ldots, 20$).
We note that the triangular lattice geometry ($p_\pm=1$) which separates DTL from ETL may lead to the complete absence of the balanced states, as we have shown in Ref.~\onlinecite{2007.02128}. 
The difference between DTL and ETL visible in \Cref{fig:all-TC_vs_p-a} is that, contrary to DTL, for ETL the characteristic temperature for given values of $p_+$ depends on the system size.
Our search for nontrivial combinations of the parameters describing the underlying network, in particular relations between the number of links $L$ and the number of triads $\Delta$, revealed that in the case of ETL the characteristic temperature approaches values common for all sizes when plotted as a function of $\bar{k} D^{5/2}$, see \Cref{fig:all-TC_vs_p-b}.

In the limit of large $p_+$ the characteristic temperature depends on $p_+$ as $T_C^U \propto p_+^{\gamma}$.
To find the exponent $\gamma$, least-squares fitting procedure was performed for those points in \Cref{fig:all-TC_vs_p} which represent graphs with $\bar{k}D^{5/2}>40$, i.e., when approximately $p_+>12.7$ for $W=10$, $p_+>22.8$ for $W=15$, and $p_+>34.5$ for $W=20$.
It reveals that $\gamma \approx 1.64\text{--}1.69$, see \Cref{tab:gamma} for the complete list of the values of $\gamma$ and their uncertainties.

\begin{table}[htbp]
 \caption{\label{tab:gamma}Exponent $\gamma$ and its uncertainty $u(\gamma)$ in $T_C^U \propto p_+^\gamma$ found for ETL's sizes 
 $W=10$, $15$, and $20$.}
\begin{ruledtabular}
 \begin{tabular}{r c c c} 
 $W$         & 10   & 15    & 20    \\
 \hline
 $\gamma$    & 1.64 & 1.644 & 1.686 \\ 
 $u(\gamma)$ & 0.02 & 0.007 & 0.006 \\
 \end{tabular}
 \end{ruledtabular}
\end{table}

\begin{figure*}[htbp]
\begin{subfigure}[b]{.48\textwidth}
\caption{\label{fig:TL-TC-and-TX--U-k3}}
\includegraphics[width=0.96\textwidth]{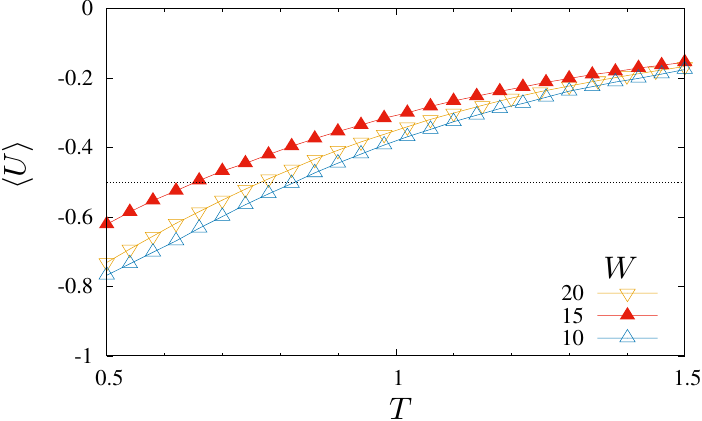}
\end{subfigure}
\begin{subfigure}[b]{.48\textwidth}
\caption{\label{fig:TL-TC-and-TX--U-kDX40}}
\includegraphics[width=0.96\textwidth]{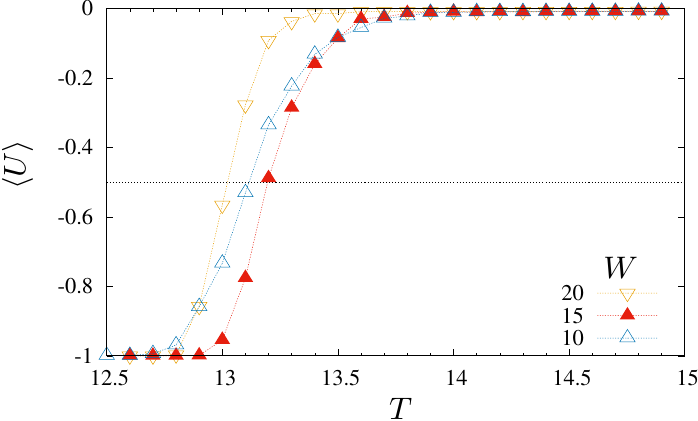}
\end{subfigure}
\begin{subfigure}[b]{.48\textwidth}
\caption{\label{fig:TL-TC-and-TX--KU-k3}}
\includegraphics[width=0.96\textwidth]{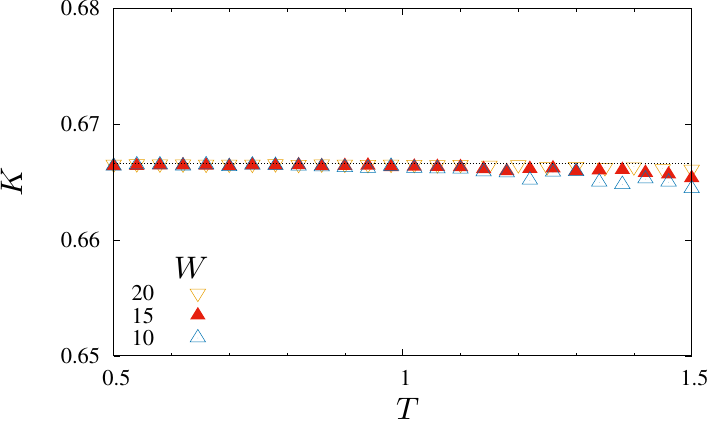}
\end{subfigure}
\begin{subfigure}[b]{.48\textwidth}
\caption{\label{fig:TL-TC-and-TX--KU-kDX40}}
\includegraphics[width=0.96\textwidth]{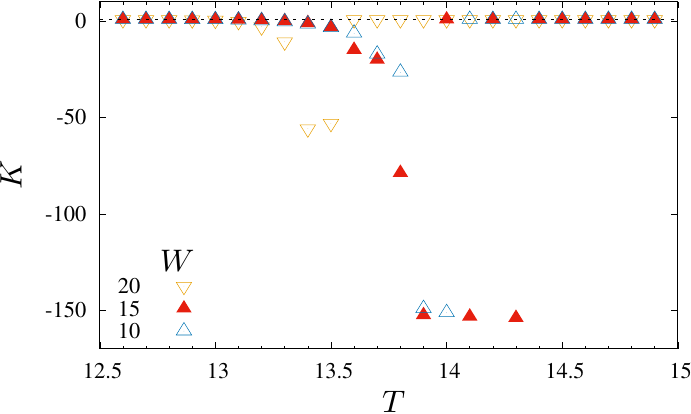}
\end{subfigure}
\begin{subfigure}[b]{.48\textwidth}
\caption{\label{fig:TL-TC-and-TX--Deltas-k3}}
\includegraphics[width=0.96\textwidth]{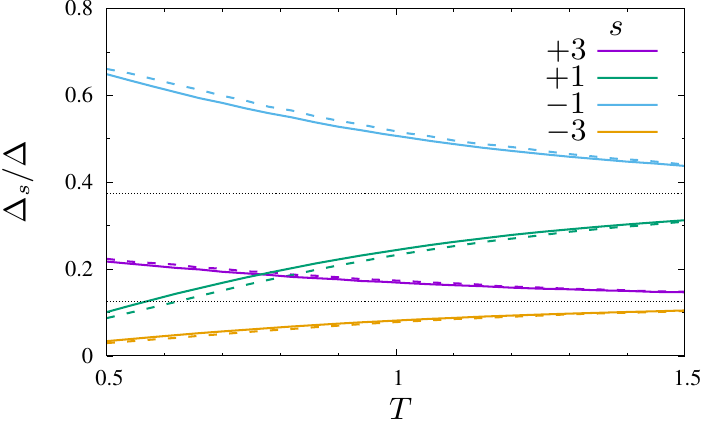}
\end{subfigure}
\begin{subfigure}[b]{.48\textwidth}
\caption{\label{fig:TL-TC-and-TX--Deltas-kDX40}}
\includegraphics[width=0.96\textwidth]{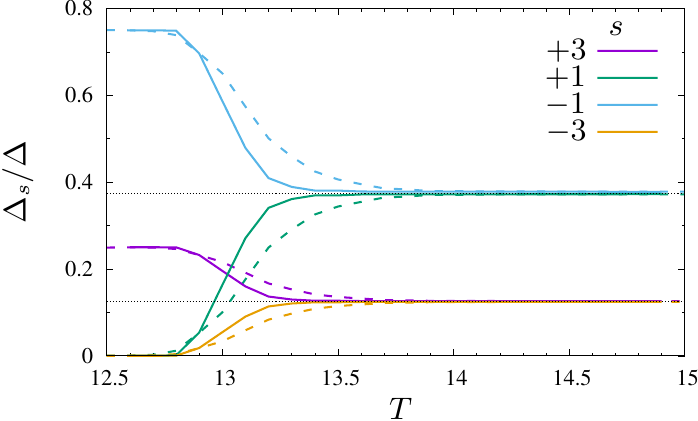}
\end{subfigure}
\begin{subfigure}[b]{.48\textwidth}
\caption{\label{fig:TL-TC-and-TX--sigmaU-k3}}
\includegraphics[width=0.96\textwidth]{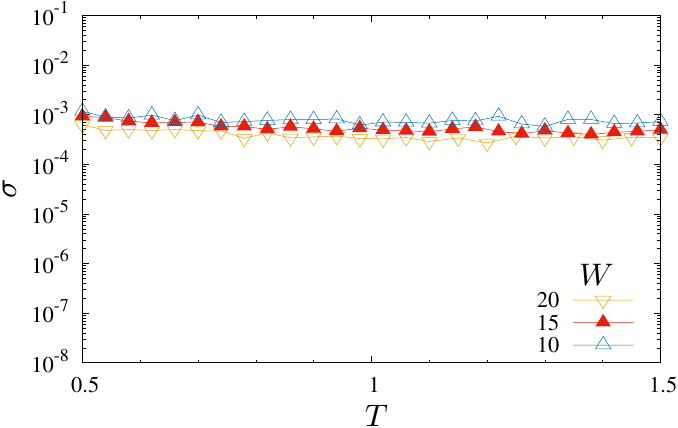}
\end{subfigure}
\begin{subfigure}[b]{.48\textwidth}
\caption{\label{fig:TL-TC-and-TX--sigmaU-kDX40}}
\includegraphics[width=0.96\textwidth]{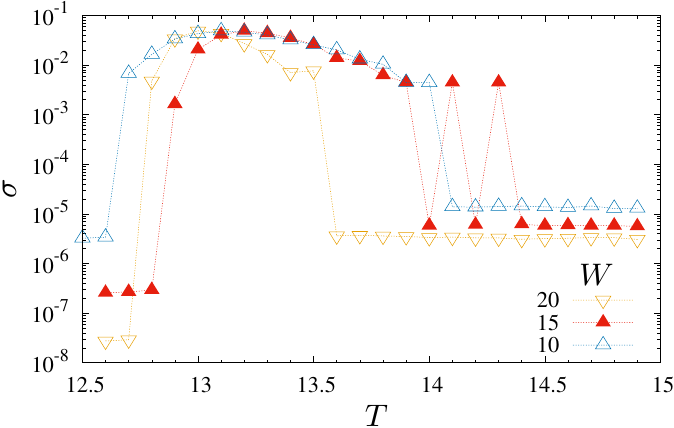}
\end{subfigure}
\caption{\label{fig:TL-TC-and-TX}(Color online). 
Left column (a, c, e, g): DTL, $p_- = 0.5$. Right column (b, d, f, h): ETL, $\bar kD^{5/2}=40$.
(a, b) Thermal evolution of the average work function $\langle U\rangle$.
(c, d) Thermal evolution of Binder cumulant $K$. Note that the presented ranges of $K$ differ significantly between (c) and (d), by several orders of magnitude. The dashed black line indicates the value of 2/3.
(e, f) Thermal evolution of fractions of all possible triads, $\Delta_s / \Delta$. Solid lines for $W=20$ and dashed lines for $W=10$.
(g, h) Thermal evolution of the standard deviation of the mean work function, $\sigma$.
In all cases, $t_{\max}=10^4$, $\tau=10^3$, $R=500$.}
\end{figure*}

The characteristic temperature $T_C^U$ found from the condition $U=-0.5$ cannot be, however, regarded as identical to the critical temperature at which a phase transition occurs.
In fact, this is not even certain that a phase transition occurs at the above discussed characteristic temperatures.
One of the possible ways to reveal the phase transition is to apply the finite size scaling analysis based on  the fourth-order Binder cumulant $K$ of the work function $U$ defined as~\cite[p. 78]{bookDL}
\begin{equation}
\label{eq:K}
K =  1 - \frac{\langle U^4 \rangle}{3 \langle U^2 \rangle^2}.
\end{equation}
The kind of thermal evolution of Binder cumulant $K(T)$ indicates the type of phase transition:
\begin{itemize}
    \item For a continuous (second order) phase transition, $K$ changes smoothly from 2/3 for low temperatures ($T\to 0$) to zero for high temperatures ($T\to\infty$) \cite[p. 508]{Binder_1997}, \cite{PhysRevE.59.218} with nontrivial behavior of $K(T_C)$ for various system sizes \cite[p. 80]{bookDL}.
    \item In the case of the first-order (abrupt) phase transition, a deep minimum of the $K(T)$ value appears in the vicinity of critical temperature $T_C$ \cite[p. 85]{bookDL}, \cite{PhysRevE.59.218}. In both limits ($T\to 0$ and $T\to\infty$) $K(T)$ tends to 2/3.
\end{itemize}

The possibility of the phase transition may be therefore verified for a particular network based on the $K(T)$ dependence.
For example, let us consider two different kinds of networks, both with sizes $W=10, 15$ and $20$: DTL with $p_-=0.5$, and ETL with $\bar{k}D^{5/2}=40$ so that size-dependence in $T_C(p)$ is circumvented.
For the first one, the characteristic temperature can be found at $T\approx 0.7\text{--}0.8$ (\Cref{fig:TL-TC-and-TX--U-k3}), but the cumulant remains almost perfectly constant, $K \approx 2/3$, in the vicinity of that characteristic temperature (\Cref{fig:TL-TC-and-TX--KU-k3}).
It means that there is no evidence of a phase transition between the balanced and imbalanced state for that DTL, and we only observe a smooth transition between the two states.
The number of triads characterized by the value of $s$, denoted as $\Delta_s$, in proportion to the total number of triads $\Delta$, is shown in \Cref{fig:TL-TC-and-TX--Deltas-k3} as a function of $T$.
It changes slowly and approaches 3/8 for the triads with $s=+1$ and $s=-1$ triads each, and 1/8 for the triads with $s=+3$ and $s=-3$ each, only when the thermal noise becomes very large, much above the characteristic temperature.
On the other hand, when we take a closer look at ETLs of various sizes, but each with $\bar{k}D^{5/2}=40$, we can see that not only the change from $\langle U \rangle = -1$ to $\langle U \rangle = 0$ is much more abrupt (\Cref{fig:TL-TC-and-TX--U-kDX40}), but more importantly there is a distinct and deep minimum in $K$ indicating the first order phase transition, as visible in \Cref{fig:TL-TC-and-TX--KU-kDX40}.
Moreover, the numbers of triads very quickly reach the limiting values when the thermal noise increases, as shown in \Cref{fig:TL-TC-and-TX--Deltas-kDX40}.
It is worth to mention that the position of $T_C^U$ (\Cref{fig:TL-TC-and-TX--U-k3,fig:TL-TC-and-TX--U-kDX40}) coincides nicely with temperature where $\Delta_{+3}=\Delta_{-1}$. However, the prevalence of $s=-1$  triads (\Cref{fig:1a}) over $s=+3$ triads (\Cref{fig:1c}) for $T>T_C^U$ cannot directly explain the crossover between balanced and imbalanced states of the system as these types of triads are balanced in Heider's sense.

Another quantity which can be used to find traces of possible phase transition is the variation of the work function, $\sigma$, which we define as the standard deviation of the mean $\langle U \rangle$ averaged over $R$ simulations.
Fluctuations in energy (work function) are proportional to the square root of the heat capacity at a constant volume. The latter value increases linearly with the size of the system, which leads to \emph{i}) the disappearance of the relative energy fluctuations $\sigma/\langle U \rangle$ inversely proportional to the size of the system and \emph{ii}) its complete disappearance in the thermodynamic limit ($W\to\infty$) regardless of the temperature. The exception to this rule is the critical point $T_C$, in which the heat capacity of the system diverges, which causes the fluctuations to be present at all size scales \cite[p. 41--42]{Binney_1992}.
As expected, $\sigma$ presented in \Cref{fig:TL-TC-and-TX--sigmaU-kDX40} increases rapidly (by at least three orders of magnitude) in the vicinity of the phase transition, while it remains virtually constant (\Cref{fig:TL-TC-and-TX--sigmaU-k3}) in the case of the smooth crossover between the imbalanced and partially balanced states in DTLs with $p_-=0.5$.

\begin{figure}[htbp]
\begin{subfigure}[b]{.48\textwidth}
\caption{\label{fig:K_vs_p_T-W10}}
\includegraphics[width=0.99\columnwidth]{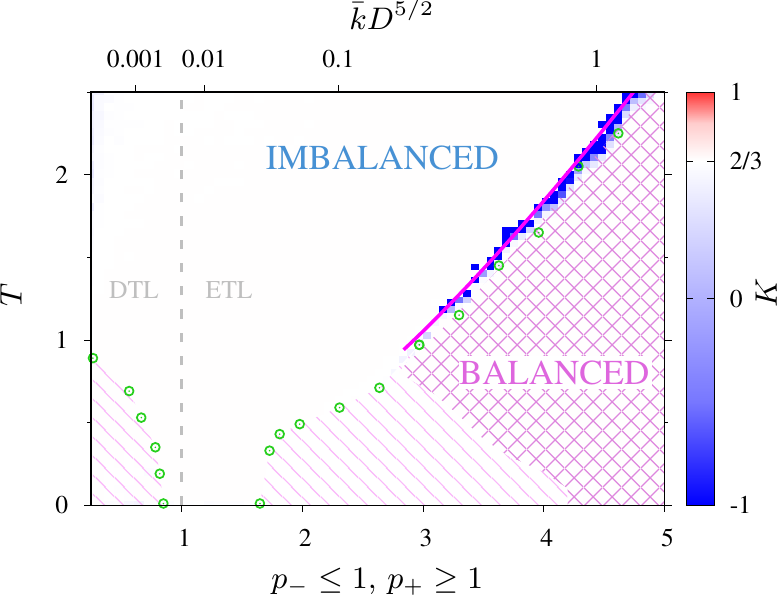}
\end{subfigure}
\begin{subfigure}[b]{.48\textwidth}
\caption{\label{fig:K_vs_p_T-W20}}
\includegraphics[width=0.99\columnwidth]{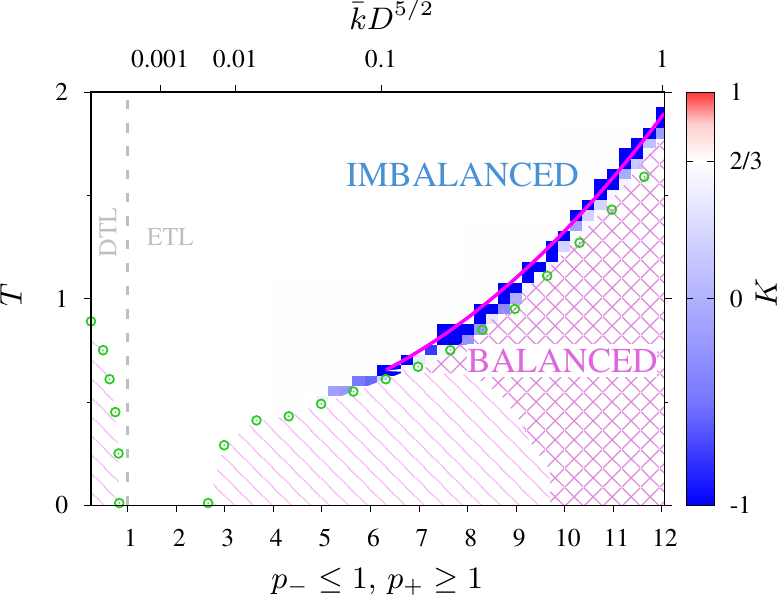}
\end{subfigure}
\caption{\label{fig:K_vs_p_T}(Color online). The fourth-order cumulant of the work function, $K(p_\pm,T)$, for (a)~$W=10$, (b)~$W=20$. Deep minimum of $K$ indicates the critical temperature $T_C^K$ (solid purple line). Green circles denote $T_C^U$ found from $\langle U \rangle=-0.5$.  Crossed lines show the balanced phase, and parallel hatching the partially balanced states.  $R=100$, $t_{\text{max}}=10^4$, $\tau=10^3$.}
\end{figure}

Finding the boundary between the phases and the points at which the phase transition occurs is therefore possible by calculating $K(p,T)$.
The results for $W=10$ and $W=20$ are shown in \Cref{fig:K_vs_p_T}.
The solid purple line in \Cref{fig:K_vs_p_T} indicates the deep minimum of $K$ and defines critical temperature $T_C^K$ which
separates the balanced and imbalanced phases, while other regions below the points corresponding to the characteristic temperatures (green circles) can be related only to the partially balanced system.
We note that $T_C^K\approx T_C^U$ for large enough values of $p_+$, where we are able to identify deep minimum of $K$.
Phase diagrams in \Cref{fig:K_vs_p_T} contain results only for relatively small values of $p$, but obviously the boundary between the two phases extends to maximum $p$ along the points defined by the characteristic temperatures (see also \Cref{fig:ETL_U_vs_p_T}).

\begin{figure*}[htbp]
\begin{subfigure}[b]{.48\textwidth}
\caption{\label{fig:Delta+3_vs_T_k}}
\includegraphics[width=0.99\textwidth]{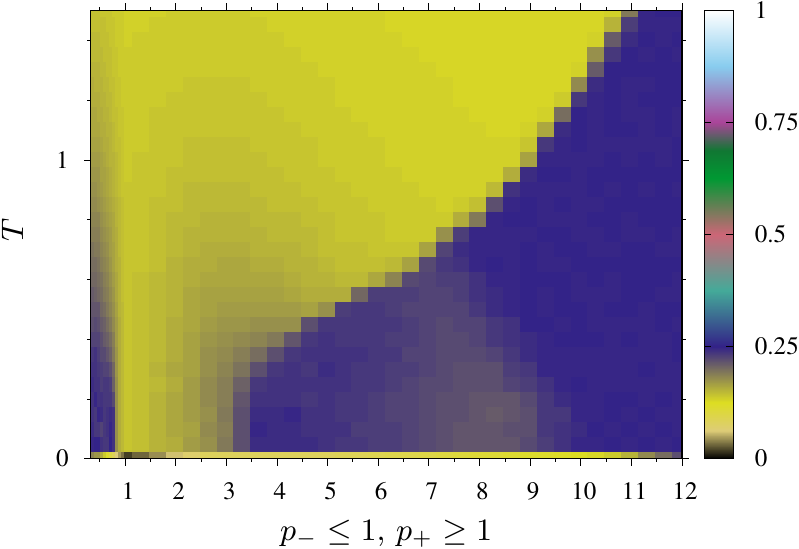}
\end{subfigure}
\begin{subfigure}[b]{.48\textwidth}
\caption{\label{fig:Delta+1_vs_T_k}}
\includegraphics[width=0.99\textwidth]{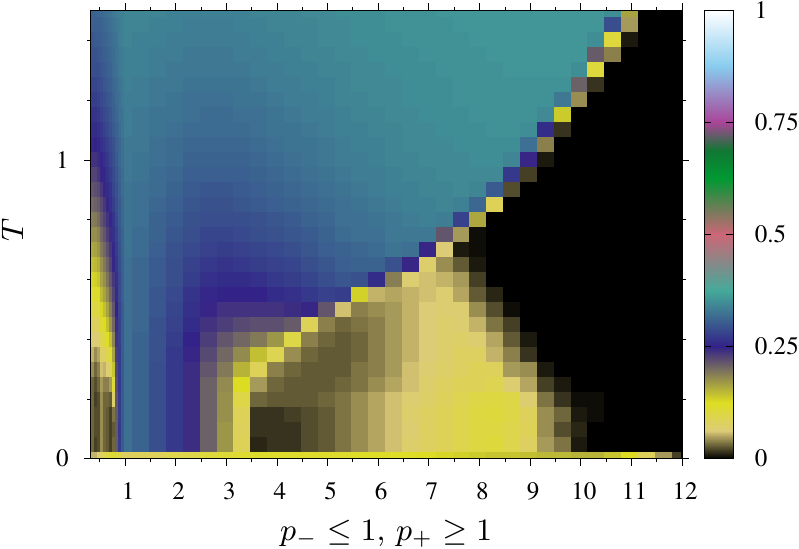}
\end{subfigure}
\begin{subfigure}[b]{.48\textwidth}
\caption{\label{fig:Delta-1_vs_T_k}}
\includegraphics[width=0.99\textwidth]{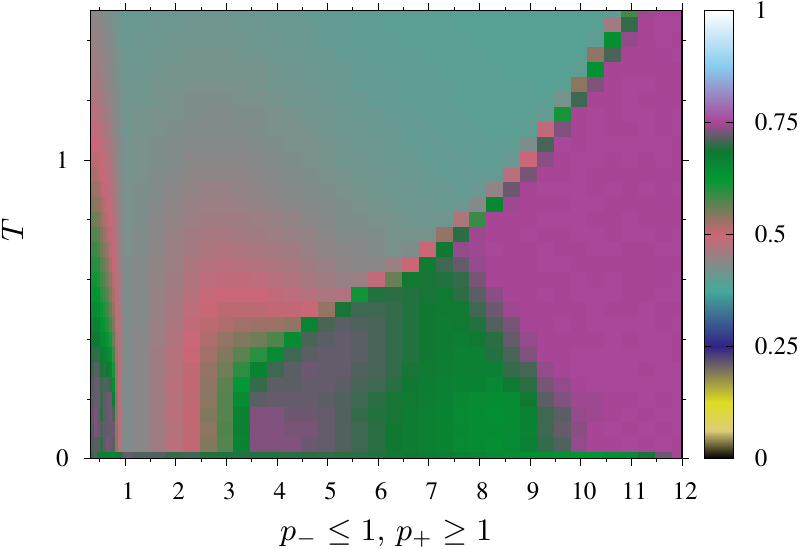}
\end{subfigure}
\begin{subfigure}[b]{.48\textwidth}
\caption{\label{fig:Delta-3_vs_T_k}}
\includegraphics[width=0.99\textwidth]{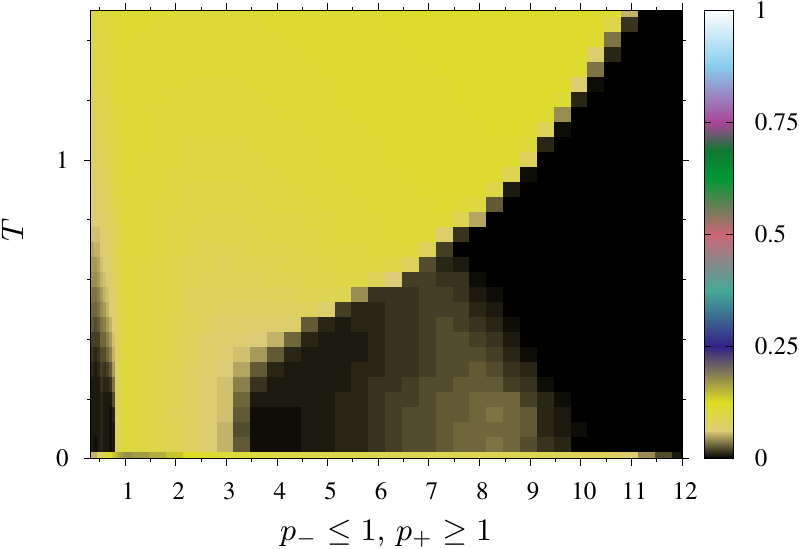}
\end{subfigure}
\caption{\label{fig:Delta_vs_T_k}(Color online). Fractions of triads, $\Delta_s(p_\pm,T)/ \Delta$, calculated for $W=20$, $R=100$, $t_{\text{max}}=10^4$, $\tau=10^3$ and all four types of triads: (a) $s=+3$, (b) $s=+1$, (c) $s=-1$, (d) $s=-3$, in DTL ($p_-<1$) and ETL ($p_+>1$). Results in the left column, (a) and (c), correspond to the balanced triads (\ref{fig:1a}, \ref{fig:1c}); in the right column, (b) and (d), to the imbalanced triads (\ref{fig:1b}, \ref{fig:1d}).}
\end{figure*}

The boundary between the balanced phase and the partially balanced states is also visible in \Cref{fig:Delta_vs_T_k} presenting the distribution of all four types of triads.
The balanced phase can be easily identified since no imbalanced triads are then observed (black areas in \Cref{fig:Delta+1_vs_T_k,fig:Delta-3_vs_T_k} for $p_+>8$), with 25\% of $s=+3$ and 75\% of $s=-1$ triads (see \Cref{fig:Delta+3_vs_T_k,fig:Delta-1_vs_T_k}).
At smaller $p_+$, roughly between 5 and 9, a partially imbalanced state is manifested below $T_C^U$ mainly by the increased number of $s=+1$ (imbalanced \ref{fig:1b}) triads at the cost of $s=-1$ (balanced \ref{fig:1c}) triads.

\section{Conclusions}

The underlying network topology has a decisive impact on the thermally driven cross-over between the balanced and imbalanced states in systems used to model the dynamics of hostile and friendly attitudes.
We illustrate that property with the example of regular triangular lattices modified by removing or adding some links.
As a result, the discussed lattices cover the full range of possible graph densities, from a very low number of links and only one or several triads up to the regular triangular lattice (DTL, $0<p_-\leq 1$), and then from the regular triangular lattice to  complete graphs (ETL, $p_+\geq 1$).

Our simulations show that reaching at least partial Heider balance is possible below the characteristic temperature $T_C^U$
if the lattice differs significantly from the triangular lattice, i.e., enough links were removed to create a DTL or added to produce an ETL.
It is related to the fact that for the ideal triangular lattice the thermal noise completely prevents even partial Heider balance~\cite{2007.02128}, which may have its origin in a large density of triads (compared to the random lattices with the same number of links), responsible for frustration similar to that observed in spin glasses.
In the case of the diluted triangular lattice, the average node degree must be reduced at least by one.
In the enhanced triangular lattices, two classes of links are present: the nearest-neighbor links preserved from the triangular lattice, and additional random links, statistically mostly long-range (which are the only kind existing in the random lattices).
The influence of those two classes of links on the existence of the balanced and imbalanced phases seems to be the opposite, with the first acting against the balanced state and the second helping to achieve it.
Finally, a sufficient number of long-range links needs to be added to overcome the effect of the triangular lattice.
The required minimum of the average node degree depends on the size of the system, which may suggest that emergence of the balanced phase demands more relatively short-ranged (but not nearest-neighbor) links, which are less likely in larger lattices when created randomly.

In DTL, the characteristic temperature
initially decreases from $T_C^U=1$ with increasing graph density and approaches zero when the average node degree $\bar{k} \approx 5$, independently on the system size.
Further increase of the graph density, beyond the regular triangular lattice, again allows to find a positive $T_C^U$ only when enough links are added, with both the required number of links and the value of the characteristic temperature now depending on the system size.
Additionally, we show that $T_C^U$ assumes values following a power law and reaches the limiting value known for complete graphs \cite{2009.10136}.

It should be noted, though, that the values of the characteristic temperature $T_C^U$ presented in this work were obtained based on the arbitrary criterion $\langle U(T) \rangle = -0.5$.
Moreover, caution must be taken even if the graph is connected but contains only a small number of noninteracting triads, which makes it difficult to define the cross-over.

Another interesting observation can be made concerning the dependence of the work function on temperature, which reveals the cross-over between balanced and imbalanced states.
That cross-over is significantly sharper at higher graph density, which resembles the situation in the case of chains of actors when the range of coupling between the nodes is larger \cite{2008.06362}.
With the help of the fourth-order Binder cumulant, we were able to show that in the case of larger graph densities, the analyzed systems undergo a phase transition of the first kind and find the critical temperatures $T_C^K$ separating the balanced and imbalanced phases.
In the limit of dense graphs, the critical temperatures given as a function of $\bar{k}D^{5/2}$ do not depend on the system size, however this intriguing property remains an open question.

In summary, our results demonstrate how the geometry of the considered lattices affects the process of reaching the structural balance in the presence of thermal noise simulated using the heat-bath dynamics.
It is shown under what conditions the balanced or partially balanced states are possible, where the phase transition of the first kind is observed, and what are the critical temperatures as functions of the bond occupation parameter.

\begin{acknowledgements}
Authors are grateful to Krzysztof Ku{\l}akowski and Zdzis{\l}aw Burda for critical reading of the manuscript and fruitful discussion.
This research was supported in part by PLGrid Infrastructure.
\end{acknowledgements}

\bibliography{this,heider,km,percolation}

\begin{thebibliography}{30}%
\makeatletter
\providecommand \@ifxundefined [1]{%
 \@ifx{#1\undefined}
}%
\providecommand \@ifnum [1]{%
 \ifnum #1\expandafter \@firstoftwo
 \else \expandafter \@secondoftwo
 \fi
}%
\providecommand \@ifx [1]{%
 \ifx #1\expandafter \@firstoftwo
 \else \expandafter \@secondoftwo
 \fi
}%
\providecommand \natexlab [1]{#1}%
\providecommand \enquote  [1]{``#1''}%
\providecommand \bibnamefont  [1]{#1}%
\providecommand \bibfnamefont [1]{#1}%
\providecommand \citenamefont [1]{#1}%
\providecommand \href@noop [0]{\@secondoftwo}%
\providecommand \href [0]{\begingroup \@sanitize@url \@href}%
\providecommand \@href[1]{\@@startlink{#1}\@@href}%
\providecommand \@@href[1]{\endgroup#1\@@endlink}%
\providecommand \@sanitize@url [0]{\catcode `\\12\catcode `\$12\catcode
  `\&12\catcode `\#12\catcode `\^12\catcode `\_12\catcode `\%12\relax}%
\providecommand \@@startlink[1]{}%
\providecommand \@@endlink[0]{}%
\providecommand \url  [0]{\begingroup\@sanitize@url \@url }%
\providecommand \@url [1]{\endgroup\@href {#1}{\urlprefix }}%
\providecommand \urlprefix  [0]{URL }%
\providecommand \Eprint [0]{\href }%
\providecommand \doibase [0]{http://dx.doi.org/}%
\providecommand \selectlanguage [0]{\@gobble}%
\providecommand \bibinfo  [0]{\@secondoftwo}%
\providecommand \bibfield  [0]{\@secondoftwo}%
\providecommand \translation [1]{[#1]}%
\providecommand \BibitemOpen [0]{}%
\providecommand \bibitemStop [0]{}%
\providecommand \bibitemNoStop [0]{.\EOS\space}%
\providecommand \EOS [0]{\spacefactor3000\relax}%
\providecommand \BibitemShut  [1]{\csname bibitem#1\endcsname}%
\let\auto@bib@innerbib\@empty
\bibitem [{\citenamefont {Heider}(1946)}]{Heider}%
  \BibitemOpen
  \bibfield  {author} {\bibinfo {author} {\bibfnamefont {F.}~\bibnamefont
  {Heider}},\ }\bibfield  {title} {\enquote {\bibinfo {title} {Attitudes and
  cognitive organization},}\ }\href {\doibase 10.1080/00223980.1946.9917275}
  {\bibfield  {journal} {\bibinfo  {journal} {The Journal of Psychology}\
  }\textbf {\bibinfo {volume} {21}},\ \bibinfo {pages} {107--112} (\bibinfo
  {year} {1946})}\BibitemShut {NoStop}%
\bibitem [{\citenamefont {Harary}(1953)}]{Harary_1953}%
  \BibitemOpen
  \bibfield  {author} {\bibinfo {author} {\bibfnamefont {F.}~\bibnamefont
  {Harary}},\ }\bibfield  {title} {\enquote {\bibinfo {title} {On the notion of
  balance of a signed graph},}\ }\href {\doibase 10.1307/mmj/1028989917}
  {\bibfield  {journal} {\bibinfo  {journal} {Michigan Mathematical Journal}\
  }\textbf {\bibinfo {volume} {2}},\ \bibinfo {pages} {143--146} (\bibinfo
  {year} {1953})}\BibitemShut {NoStop}%
\bibitem [{\citenamefont {Cartwright}\ and\ \citenamefont
  {Harary}(1956)}]{Cartwright_1956}%
  \BibitemOpen
  \bibfield  {author} {\bibinfo {author} {\bibfnamefont {D.}~\bibnamefont
  {Cartwright}}\ and\ \bibinfo {author} {\bibfnamefont {F.}~\bibnamefont
  {Harary}},\ }\bibfield  {title} {\enquote {\bibinfo {title} {Structural
  balance: {A} generalization of {H}eider's theory},}\ }\href {\doibase
  10.1037/h0046049} {\bibfield  {journal} {\bibinfo  {journal} {Psychological
  Review}\ }\textbf {\bibinfo {volume} {63}},\ \bibinfo {pages} {277–293}
  (\bibinfo {year} {1956})}\BibitemShut {NoStop}%
\bibitem [{\citenamefont {Harary}(1959)}]{Harary_1959}%
  \BibitemOpen
  \bibfield  {author} {\bibinfo {author} {\bibfnamefont {F.}~\bibnamefont
  {Harary}},\ }\bibfield  {title} {\enquote {\bibinfo {title} {On the
  measurement of structural balance},}\ }\href {\doibase 10.1002/bs.3830040405}
  {\bibfield  {journal} {\bibinfo  {journal} {Behavioral Science}\ }\textbf
  {\bibinfo {volume} {4}},\ \bibinfo {pages} {316--323} (\bibinfo {year}
  {1959})}\BibitemShut {NoStop}%
\bibitem [{\citenamefont {Davis}(1967)}]{Davis_1967}%
  \BibitemOpen
  \bibfield  {author} {\bibinfo {author} {\bibfnamefont {J.~A.}\ \bibnamefont
  {Davis}},\ }\bibfield  {title} {\enquote {\bibinfo {title} {Clustering and
  structural balance in graphs},}\ }\href {\doibase 10.1177/001872676702000206}
  {\bibfield  {journal} {\bibinfo  {journal} {Human Relations}\ }\textbf
  {\bibinfo {volume} {20}},\ \bibinfo {pages} {181--187} (\bibinfo {year}
  {1967})}\BibitemShut {NoStop}%
\bibitem [{\citenamefont {Harary}\ \emph {et~al.}(1965)\citenamefont {Harary},
  \citenamefont {Norman},\ and\ \citenamefont {Cartwright}}]{Harary}%
  \BibitemOpen
  \bibfield  {author} {\bibinfo {author} {\bibfnamefont {F.}~\bibnamefont
  {Harary}}, \bibinfo {author} {\bibfnamefont {R.~Z.}\ \bibnamefont {Norman}},
  \ and\ \bibinfo {author} {\bibfnamefont {D.}~\bibnamefont {Cartwright}},\
  }\href@noop {} {\emph {\bibinfo {title} {Structural Models: {A}n Introduction
  to the Theory of Directed Graphs}}},\ \bibinfo {edition} {3rd}\ ed.\
  (\bibinfo  {publisher} {John Wiley and Sons},\ \bibinfo {address} {New
  York},\ \bibinfo {year} {1965})\BibitemShut {NoStop}%
\bibitem [{\citenamefont {Malarz}\ and\ \citenamefont
  {Ku{\l}akowski}(2021{\natexlab{a}})}]{2008.06362}%
  \BibitemOpen
  \bibfield  {author} {\bibinfo {author} {\bibfnamefont {K.}~\bibnamefont
  {Malarz}}\ and\ \bibinfo {author} {\bibfnamefont {K.}~\bibnamefont
  {Ku{\l}akowski}},\ }\bibfield  {title} {\enquote {\bibinfo {title} {Heider
  balance of a chain of actors as dependent on the interaction range and a
  thermal noise},}\ }\href {\doibase 10.1016/j.physa.2020.125640} {\bibfield
  {journal} {\bibinfo  {journal} {Physica A}\ }\textbf {\bibinfo {volume}
  {567}},\ \bibinfo {pages} {125640} (\bibinfo {year}
  {2021}{\natexlab{a}})}\BibitemShut {NoStop}%
\bibitem [{\citenamefont {Antal}\ \emph {et~al.}(2005)\citenamefont {Antal},
  \citenamefont {Krapivsky},\ and\ \citenamefont {Redner}}]{Antal_2005}%
  \BibitemOpen
  \bibfield  {author} {\bibinfo {author} {\bibfnamefont {T.}~\bibnamefont
  {Antal}}, \bibinfo {author} {\bibfnamefont {P.~L.}\ \bibnamefont
  {Krapivsky}}, \ and\ \bibinfo {author} {\bibfnamefont {S.}~\bibnamefont
  {Redner}},\ }\bibfield  {title} {\enquote {\bibinfo {title} {Dynamics of
  social balance on networks},}\ }\href {\doibase 10.1103/PhysRevE.72.036121}
  {\bibfield  {journal} {\bibinfo  {journal} {Physical Review E}\ }\textbf
  {\bibinfo {volume} {72}},\ \bibinfo {pages} {036121} (\bibinfo {year}
  {2005})}\BibitemShut {NoStop}%
\bibitem [{\citenamefont {Ku{\l}akowski}(2007)}]{ISI:000247470200019}%
  \BibitemOpen
  \bibfield  {author} {\bibinfo {author} {\bibfnamefont {K.}~\bibnamefont
  {Ku{\l}akowski}},\ }\bibfield  {title} {\enquote {\bibinfo {title} {Some
  recent attempts to simulate the {H}eider balance problem},}\ }\href {\doibase
  10.1109/MCSE.2007.85} {\bibfield  {journal} {\bibinfo  {journal} {Computing
  in Science \& Engineering}\ }\textbf {\bibinfo {volume} {9}},\ \bibinfo
  {pages} {80--85} (\bibinfo {year} {2007})}\BibitemShut {NoStop}%
\bibitem [{\citenamefont {Belaza}\ \emph {et~al.}(2017)\citenamefont {Belaza},
  \citenamefont {Hoefman}, \citenamefont {Ryckebusch}, \citenamefont {Bramson},
  \citenamefont {van~den Heuvel},\ and\ \citenamefont {Schoors}}]{Belaza_2017}%
  \BibitemOpen
  \bibfield  {author} {\bibinfo {author} {\bibfnamefont {A.~M.}\ \bibnamefont
  {Belaza}}, \bibinfo {author} {\bibfnamefont {K.}~\bibnamefont {Hoefman}},
  \bibinfo {author} {\bibfnamefont {J.}~\bibnamefont {Ryckebusch}}, \bibinfo
  {author} {\bibfnamefont {A.}~\bibnamefont {Bramson}}, \bibinfo {author}
  {\bibfnamefont {M.}~\bibnamefont {van~den Heuvel}}, \ and\ \bibinfo {author}
  {\bibfnamefont {K.}~\bibnamefont {Schoors}},\ }\bibfield  {title} {\enquote
  {\bibinfo {title} {Statistical physics of balance theory},}\ }\href {\doibase
  10.1371/journal.pone.0183696} {\bibfield  {journal} {\bibinfo  {journal}
  {Plos One}\ }\textbf {\bibinfo {volume} {12}},\ \bibinfo {pages} {e0183696}
  (\bibinfo {year} {2017})}\BibitemShut {NoStop}%
\bibitem [{\citenamefont {Masoumi}\ \emph {et~al.}(2021)\citenamefont
  {Masoumi}, \citenamefont {Oloomi}, \citenamefont {Kargaran}, \citenamefont
  {Hosseiny},\ and\ \citenamefont {Jafari}}]{2008.00537}%
  \BibitemOpen
  \bibfield  {author} {\bibinfo {author} {\bibfnamefont {R.}~\bibnamefont
  {Masoumi}}, \bibinfo {author} {\bibfnamefont {F.}~\bibnamefont {Oloomi}},
  \bibinfo {author} {\bibfnamefont {A.}~\bibnamefont {Kargaran}}, \bibinfo
  {author} {\bibfnamefont {A.}~\bibnamefont {Hosseiny}}, \ and\ \bibinfo
  {author} {\bibfnamefont {G.~R.}\ \bibnamefont {Jafari}},\ }\bibfield  {title}
  {\enquote {\bibinfo {title} {Mean-field solution for critical behavior of
  signed networks in competitive balance theory},}\ }\href {\doibase
  10.1103/PhysRevE.103.052301} {\bibfield  {journal} {\bibinfo  {journal}
  {Physical Review E}\ }\textbf {\bibinfo {volume} {103}},\ \bibinfo {pages}
  {052301} (\bibinfo {year} {2021})}\BibitemShut {NoStop}%
\bibitem [{\citenamefont {Kargaran}\ and\ \citenamefont
  {Jafari}(2021)}]{2010.10036}%
  \BibitemOpen
  \bibfield  {author} {\bibinfo {author} {\bibfnamefont {A.}~\bibnamefont
  {Kargaran}}\ and\ \bibinfo {author} {\bibfnamefont {G.~R.}\ \bibnamefont
  {Jafari}},\ }\bibfield  {title} {\enquote {\bibinfo {title} {Heider and
  coevolutionary balance: {F}rom discrete to continuous phase transition},}\
  }\href {\doibase 10.1103/PhysRevE.103.052302} {\bibfield  {journal} {\bibinfo
   {journal} {Physical Review E}\ }\textbf {\bibinfo {volume} {103}},\ \bibinfo
  {pages} {052302} (\bibinfo {year} {2021})}\BibitemShut {NoStop}%
\bibitem [{\citenamefont {Minh~Pham}\ \emph {et~al.}(2020)\citenamefont
  {Minh~Pham}, \citenamefont {Kondor}, \citenamefont {Hanel},\ and\
  \citenamefont {Thurner}}]{Pham_2005.01815}%
  \BibitemOpen
  \bibfield  {author} {\bibinfo {author} {\bibfnamefont {T.}~\bibnamefont
  {Minh~Pham}}, \bibinfo {author} {\bibfnamefont {I.}~\bibnamefont {Kondor}},
  \bibinfo {author} {\bibfnamefont {R.}~\bibnamefont {Hanel}}, \ and\ \bibinfo
  {author} {\bibfnamefont {S.}~\bibnamefont {Thurner}},\ }\bibfield  {title}
  {\enquote {\bibinfo {title} {The effect of social balance on social
  fragmentation},}\ }\href {\doibase 10.1098/rsif.2020.0752} {\bibfield
  {journal} {\bibinfo  {journal} {Journal of The Royal Society Interface}\
  }\textbf {\bibinfo {volume} {17}},\ \bibinfo {pages} {20200752} (\bibinfo
  {year} {2020})}\BibitemShut {NoStop}%
\bibitem [{\citenamefont {Oishi}\ \emph {et~al.}(2021)\citenamefont {Oishi},
  \citenamefont {Miyano}, \citenamefont {Kaski},\ and\ \citenamefont
  {Shimada}}]{2104.10568}%
  \BibitemOpen
  \bibfield  {author} {\bibinfo {author} {\bibfnamefont {K.}~\bibnamefont
  {Oishi}}, \bibinfo {author} {\bibfnamefont {S.}~\bibnamefont {Miyano}},
  \bibinfo {author} {\bibfnamefont {K.}~\bibnamefont {Kaski}}, \ and\ \bibinfo
  {author} {\bibfnamefont {T.}~\bibnamefont {Shimada}},\ }\bibfield  {title}
  {\enquote {\bibinfo {title} {Balanced-imbalanced transitions in indirect
  reciprocity dynamics on networks},}\ }\href {\doibase
  10.1103/PhysRevE.104.024310} {\bibfield  {journal} {\bibinfo  {journal}
  {Physical Review E}\ }\textbf {\bibinfo {volume} {104}},\ \bibinfo {pages}
  {024310} (\bibinfo {year} {2021})}\BibitemShut {NoStop}%
\bibitem [{\citenamefont {Rabbani}\ \emph {et~al.}(2019)\citenamefont
  {Rabbani}, \citenamefont {Shirazi},\ and\ \citenamefont
  {Jafari}}]{PhysRevE.99.062302}%
  \BibitemOpen
  \bibfield  {author} {\bibinfo {author} {\bibfnamefont {F.}~\bibnamefont
  {Rabbani}}, \bibinfo {author} {\bibfnamefont {A.~H.}\ \bibnamefont
  {Shirazi}}, \ and\ \bibinfo {author} {\bibfnamefont {G.~R.}\ \bibnamefont
  {Jafari}},\ }\bibfield  {title} {\enquote {\bibinfo {title} {Mean-field
  solution of structural balance dynamics in nonzero temperature},}\ }\href
  {\doibase 10.1103/PhysRevE.99.062302} {\bibfield  {journal} {\bibinfo
  {journal} {Physical Review E}\ }\textbf {\bibinfo {volume} {99}},\ \bibinfo
  {pages} {062302} (\bibinfo {year} {2019})}\BibitemShut {NoStop}%
\bibitem [{\citenamefont {Malarz}\ and\ \citenamefont
  {Ho{\l}yst}(2019)}]{1911.13048}%
  \BibitemOpen
  \bibfield  {author} {\bibinfo {author} {\bibfnamefont {K.}~\bibnamefont
  {Malarz}}\ and\ \bibinfo {author} {\bibfnamefont {J.~A.}\ \bibnamefont
  {Ho{\l}yst}},\ }\href@noop {} {\enquote {\bibinfo {title} {Comment on
  `{M}ean-field solution of structural balance dynamics in nonzero
  temperature'},}\ } (\bibinfo {year} {2019}),\ \Eprint
  {http://arxiv.org/abs/1911.13048} {arXiv:1911.13048 [physics.soc-ph]}
  \BibitemShut {NoStop}%
\bibitem [{\citenamefont {Shojaei}\ \emph {et~al.}(2019)\citenamefont
  {Shojaei}, \citenamefont {Manshour},\ and\ \citenamefont
  {Montakhab}}]{PhysRevE.100.022303}%
  \BibitemOpen
  \bibfield  {author} {\bibinfo {author} {\bibfnamefont {R.}~\bibnamefont
  {Shojaei}}, \bibinfo {author} {\bibfnamefont {P.}~\bibnamefont {Manshour}}, \
  and\ \bibinfo {author} {\bibfnamefont {A.}~\bibnamefont {Montakhab}},\
  }\bibfield  {title} {\enquote {\bibinfo {title} {Phase transition in a
  network model of social balance with {G}lauber dynamics},}\ }\href {\doibase
  10.1103/PhysRevE.100.022303} {\bibfield  {journal} {\bibinfo  {journal}
  {Physical Review E}\ }\textbf {\bibinfo {volume} {100}},\ \bibinfo {pages}
  {022303} (\bibinfo {year} {2019})}\BibitemShut {NoStop}%
\bibitem [{\citenamefont {Malarz}\ and\ \citenamefont
  {Ku{\l}akowski}(2021{\natexlab{b}})}]{2009.10136}%
  \BibitemOpen
  \bibfield  {author} {\bibinfo {author} {\bibfnamefont {K.}~\bibnamefont
  {Malarz}}\ and\ \bibinfo {author} {\bibfnamefont {K.}~\bibnamefont
  {Ku{\l}akowski}},\ }\bibfield  {title} {\enquote {\bibinfo {title} {Comment
  on `{P}hase transition in a network model of social balance with {G}lauber
  dynamics'},}\ }\href {\doibase 10.1103/PhysRevE.103.066301} {\bibfield
  {journal} {\bibinfo  {journal} {Physical Review E}\ }\textbf {\bibinfo
  {volume} {103}},\ \bibinfo {pages} {066301} (\bibinfo {year}
  {2021}{\natexlab{b}})}\BibitemShut {NoStop}%
\bibitem [{\citenamefont {Manshour}\ and\ \citenamefont
  {Montakhab}(2021)}]{2011.07501}%
  \BibitemOpen
  \bibfield  {author} {\bibinfo {author} {\bibfnamefont {P.}~\bibnamefont
  {Manshour}}\ and\ \bibinfo {author} {\bibfnamefont {A.}~\bibnamefont
  {Montakhab}},\ }\bibfield  {title} {\enquote {\bibinfo {title} {Reply to
  ``{C}omment on `{P}hase transition in a network model of social balance with
  {G}lauber dynamics'"},}\ }\href {\doibase 10.1103/PhysRevE.103.066302}
  {\bibfield  {journal} {\bibinfo  {journal} {Physical Review E}\ }\textbf
  {\bibinfo {volume} {103}},\ \bibinfo {pages} {066302} (\bibinfo {year}
  {2021})}\BibitemShut {NoStop}%
\bibitem [{\citenamefont {Malarz}\ \emph {et~al.}(2020)\citenamefont {Malarz},
  \citenamefont {Wo{\l}oszyn},\ and\ \citenamefont
  {Ku{\l}akowski}}]{2005.11402}%
  \BibitemOpen
  \bibfield  {author} {\bibinfo {author} {\bibfnamefont {K.}~\bibnamefont
  {Malarz}}, \bibinfo {author} {\bibfnamefont {M.}~\bibnamefont {Wo{\l}oszyn}},
  \ and\ \bibinfo {author} {\bibfnamefont {K.}~\bibnamefont {Ku{\l}akowski}},\
  }\bibfield  {title} {\enquote {\bibinfo {title} {Towards the {H}eider balance
  with a cellular automaton},}\ }\href {\doibase 10.1016/j.physd.2020.132556}
  {\bibfield  {journal} {\bibinfo  {journal} {Physica D}\ }\textbf {\bibinfo
  {volume} {411}},\ \bibinfo {pages} {132556} (\bibinfo {year}
  {2020})}\BibitemShut {NoStop}%
\bibitem [{\citenamefont {Malarz}\ and\ \citenamefont
  {Wo{\l}oszyn}(2020)}]{2007.02128}%
  \BibitemOpen
  \bibfield  {author} {\bibinfo {author} {\bibfnamefont {K.}~\bibnamefont
  {Malarz}}\ and\ \bibinfo {author} {\bibfnamefont {M.}~\bibnamefont
  {Wo{\l}oszyn}},\ }\bibfield  {title} {\enquote {\bibinfo {title} {Expulsion
  from structurally balanced paradise},}\ }\href {\doibase 10.1063/5.0022922}
  {\bibfield  {journal} {\bibinfo  {journal} {Chaos}\ }\textbf {\bibinfo
  {volume} {30}},\ \bibinfo {pages} {121103} (\bibinfo {year}
  {2020})}\BibitemShut {NoStop}%
\bibitem [{\citenamefont {G\'orski}\ \emph {et~al.}(2020)\citenamefont
  {G\'orski}, \citenamefont {Bochenina}, \citenamefont {Ho\l{}yst},\ and\
  \citenamefont {D'Souza}}]{PhysRevLett.125.078302}%
  \BibitemOpen
  \bibfield  {author} {\bibinfo {author} {\bibfnamefont {P.~J.}\ \bibnamefont
  {G\'orski}}, \bibinfo {author} {\bibfnamefont {K.}~\bibnamefont {Bochenina}},
  \bibinfo {author} {\bibfnamefont {J.~A.}\ \bibnamefont {Ho\l{}yst}}, \ and\
  \bibinfo {author} {\bibfnamefont {R.~M.}\ \bibnamefont {D'Souza}},\
  }\bibfield  {title} {\enquote {\bibinfo {title} {Homophily based on few
  attributes can impede structural balance},}\ }\href {\doibase
  10.1103/PhysRevLett.125.078302} {\bibfield  {journal} {\bibinfo  {journal}
  {Physical Review Letters}\ }\textbf {\bibinfo {volume} {125}},\ \bibinfo
  {pages} {078302} (\bibinfo {year} {2020})}\BibitemShut {NoStop}%
\bibitem [{\citenamefont {Krawczyk}\ \emph {et~al.}(2017)\citenamefont
  {Krawczyk}, \citenamefont {Kaluzny},\ and\ \citenamefont
  {Ku{\l}akowski}}]{Krawczyk_2017}%
  \BibitemOpen
  \bibfield  {author} {\bibinfo {author} {\bibfnamefont {M.~J.}\ \bibnamefont
  {Krawczyk}}, \bibinfo {author} {\bibfnamefont {S.}~\bibnamefont {Kaluzny}}, \
  and\ \bibinfo {author} {\bibfnamefont {K.}~\bibnamefont {Ku{\l}akowski}},\
  }\bibfield  {title} {\enquote {\bibinfo {title} {A small chance of
  paradise---{E}quivalence of balanced states},}\ }\href {\doibase
  10.1209/0295-5075/118/58005} {\bibfield  {journal} {\bibinfo  {journal}
  {{EPL} (Europhysics Letters)}\ }\textbf {\bibinfo {volume} {118}},\ \bibinfo
  {pages} {58005} (\bibinfo {year} {2017})}\BibitemShut {NoStop}%
\bibitem [{\citenamefont {Michael}(1986)}]{PhysRevB.33.7861}%
  \BibitemOpen
  \bibfield  {author} {\bibinfo {author} {\bibfnamefont {C.}~\bibnamefont
  {Michael}},\ }\bibfield  {title} {\enquote {\bibinfo {title} {Fast heat-bath
  algorithm for the {I}sing model},}\ }\href {\doibase
  10.1103/PhysRevB.33.7861} {\bibfield  {journal} {\bibinfo  {journal}
  {Physical Review B}\ }\textbf {\bibinfo {volume} {33}},\ \bibinfo {pages}
  {7861--7862} (\bibinfo {year} {1986})}\BibitemShut {NoStop}%
\bibitem [{\citenamefont {Loison}\ \emph {et~al.}(2004)\citenamefont {Loison},
  \citenamefont {Qin}, \citenamefont {Schotte},\ and\ \citenamefont
  {Jin}}]{Loison_2004}%
  \BibitemOpen
  \bibfield  {author} {\bibinfo {author} {\bibfnamefont {D.}~\bibnamefont
  {Loison}}, \bibinfo {author} {\bibfnamefont {C.~L.}\ \bibnamefont {Qin}},
  \bibinfo {author} {\bibfnamefont {K.~D.}\ \bibnamefont {Schotte}}, \ and\
  \bibinfo {author} {\bibfnamefont {X.~F.}\ \bibnamefont {Jin}},\ }\bibfield
  {title} {\enquote {\bibinfo {title} {Canonical local algorithms for spin
  systems: heat bath and {H}asting’s methods},}\ }\href {\doibase
  10.1140/epjb/e2004-00332-5} {\bibfield  {journal} {\bibinfo  {journal} {The
  European Physical Journal B}\ }\textbf {\bibinfo {volume} {41}},\ \bibinfo
  {pages} {395--412} (\bibinfo {year} {2004})}\BibitemShut {NoStop}%
\bibitem [{\citenamefont {Stauffer}\ and\ \citenamefont
  {Aharony}(1994)}]{bookDS}%
  \BibitemOpen
  \bibfield  {author} {\bibinfo {author} {\bibfnamefont {D.}~\bibnamefont
  {Stauffer}}\ and\ \bibinfo {author} {\bibfnamefont {A.}~\bibnamefont
  {Aharony}},\ }\href {\doibase 10.1201/9781315274386} {\emph {\bibinfo {title}
  {Introduction to Percolation Theory}}},\ \bibinfo {edition} {2nd}\ ed.\
  (\bibinfo  {publisher} {Taylor and Francis},\ \bibinfo {address} {London},\
  \bibinfo {year} {1994})\BibitemShut {NoStop}%
\bibitem [{\citenamefont {Landau}\ and\ \citenamefont {Binder}(2005)}]{bookDL}%
  \BibitemOpen
  \bibfield  {author} {\bibinfo {author} {\bibfnamefont {D.~P.}\ \bibnamefont
  {Landau}}\ and\ \bibinfo {author} {\bibfnamefont {K.}~\bibnamefont
  {Binder}},\ }\href {\doibase 10.1017/CBO9780511614460} {\emph {\bibinfo
  {title} {A Guide to Monte Carlo Simulations in Statistical Physics}}},\
  \bibinfo {edition} {2nd}\ ed.\ (\bibinfo  {publisher} {Cambridge UP},\
  \bibinfo {address} {Cambridge},\ \bibinfo {year} {2005})\BibitemShut
  {NoStop}%
\bibitem [{\citenamefont {Binder}(1997)}]{Binder_1997}%
  \BibitemOpen
  \bibfield  {author} {\bibinfo {author} {\bibfnamefont {K.}~\bibnamefont
  {Binder}},\ }\bibfield  {title} {\enquote {\bibinfo {title} {Applications of
  {M}onte {C}arlo methods to statistical physics},}\ }\href {\doibase
  10.1088/0034-4885/60/5/001} {\bibfield  {journal} {\bibinfo  {journal}
  {Reports on Progress in Physics}\ }\textbf {\bibinfo {volume} {60}},\
  \bibinfo {pages} {487--559} (\bibinfo {year} {1997})}\BibitemShut {NoStop}%
\bibitem [{\citenamefont {Acharyya}(1999)}]{PhysRevE.59.218}%
  \BibitemOpen
  \bibfield  {author} {\bibinfo {author} {\bibfnamefont {M.}~\bibnamefont
  {Acharyya}},\ }\bibfield  {title} {\enquote {\bibinfo {title} {Nonequilibrium
  phase transition in the kinetic {I}sing model: {E}xistence of a tricritical
  point and stochastic resonance},}\ }\href {\doibase 10.1103/PhysRevE.59.218}
  {\bibfield  {journal} {\bibinfo  {journal} {Physical Review E}\ }\textbf
  {\bibinfo {volume} {59}},\ \bibinfo {pages} {218--221} (\bibinfo {year}
  {1999})}\BibitemShut {NoStop}%
\bibitem [{\citenamefont {Binney}\ \emph {et~al.}(1992)\citenamefont {Binney},
  \citenamefont {Dowrick}, \citenamefont {Fisher},\ and\ \citenamefont
  {Newman}}]{Binney_1992}%
  \BibitemOpen
  \bibfield  {author} {\bibinfo {author} {\bibfnamefont {J.~J.}\ \bibnamefont
  {Binney}}, \bibinfo {author} {\bibfnamefont {N.~J.}\ \bibnamefont {Dowrick}},
  \bibinfo {author} {\bibfnamefont {A.~J.}\ \bibnamefont {Fisher}}, \ and\
  \bibinfo {author} {\bibfnamefont {M.~E.~J.}\ \bibnamefont {Newman}},\
  }\href@noop {} {\emph {\bibinfo {title} {The Theory of Critical Phenomena:
  {A}n Introduction to the Renormalization Group}}}\ (\bibinfo  {publisher}
  {Oxford University Press, Inc.},\ \bibinfo {address} {New York},\ \bibinfo
  {year} {1992})\BibitemShut {NoStop}%
\end{thebibliography}%

\end{document}